\documentclass[11pt]{article}
\pdfoutput = 1
\usepackage[utf8]{inputenc}

\usepackage[margin = 2.2cm]{geometry}
\usepackage{color}
\definecolor{darkgreen}{rgb}{0,0.5,0}
\definecolor{darkblue}{rgb}{0,0,0.6}
\definecolor{purple}{rgb}{0.4,.2,0.7}

\usepackage[bookmarks = false, colorlinks = true, linkcolor = blue, citecolor = purple]{hyperref}

\usepackage{aas_macros,empheq,fancybox,graphicx,multirow}
\usepackage[bottom]{footmisc}
\usepackage{aas_macros}
\usepackage{esint}

\usepackage{amsfonts}
\usepackage{amsmath}
\usepackage{amssymb}
\usepackage{appendix}
\usepackage{cite}
\usepackage{enumitem}
\usepackage[mathcal]{eucal}
\usepackage{float}
\usepackage{graphicx}
\usepackage{mathabx}
\usepackage{mathrsfs}
\usepackage{pifont}
\usepackage{setspace}
\usepackage{slashed}
\usepackage{titlesec}
\usepackage{subcaption}
\usepackage{comment}
\usepackage{wrapfig}

\renewcommand{\i}{\mathrm{i}}

\DeclareMathOperator{\tr}{Tr}
\DeclareMathOperator{\Tr}{Tr}
\DeclareMathOperator{\vol}{vol}

\renewcommand{\d}{\mathrm{d}}
\newcommand{\pin}{\text{Pin}^+}
\newcommand{\ztwotwo}{\mathbb{Z}_2\times \mathbb{Z}_2}

\begin{document}

\thispagestyle{empty}

\begin{center}
    ~
    \vskip10mm

     {\LARGE  {\textsc{Matrix ensembles with global symmetries \\ \vspace{5pt} and 't Hooft anomalies from 2d gauge theory}}}
    \vskip10mm
    
    Daniel Kapec,$^{a}$  Raghu Mahajan,$^{a,b}$ and Douglas Stanford$^{a,c}$\\
    \vskip1em
    {\it
        $^a$ School of Natural Sciences, Institute for Advanced Study, Princeton, NJ 08540, USA\\ \vskip1mm
        $^b$  Department of Physics, Princeton University, Princeton,  NJ 08540, USA\\ \vskip1mm
        $^c$ Department of Physics, Stanford University, Stanford, CA 94305, USA\\ \vskip1mm
    }
    \vskip5mm
    \tt{kapec@ias.edu, raghu\_m@princeton.edu, salguod@stanford.edu}
\end{center}
\vspace{10mm}

\begin{abstract}
\noindent
The Hilbert space of a quantum system with internal global symmetry $G$ decomposes into sectors labelled by  irreducible representations of $G$. 
If the system is chaotic, the energies in each sector should separately resemble ordinary random matrix theory.
We show that such ``sector-wise'' random matrix ensembles arise as the boundary dual of two-dimensional gravity with a $G$ gauge field in the bulk. 
Within each sector, the eigenvalue density is enhanced by a nontrivial factor of the dimension of the representation, and the ground state energy is determined by the quadratic Casimir. 
We study the consequences of 't Hooft anomalies in the matrix ensembles, which are incorporated by adding specific topological terms to the gauge theory action. 
The effect is to introduce projective representations into the decomposition of the Hilbert space.
Finally, we consider ensembles with $G$ symmetry and time reversal symmetry, and analyze a simple case of a mixed anomaly between time reversal and an internal $\mathbb{Z}_2$ symmetry.

\end{abstract}
\pagebreak


\setcounter{tocdepth}{2}
{\hypersetup{linkcolor=black}
\tableofcontents}

\section{Introduction}
Two-dimensional quantum gravity arises in several different physical contexts. 
It serves as the simplest toy model of higher-dimensional quantum gravity, as the worldsheet description of the first-quantized string, and also as an approximation to the dynamics in the throats of higher-dimensional near-extremal black holes. 
The most successful approach to these models exploits an old idea \cite{tHooft:1973alw,Brezin:1977sv} which sought to model the integral over random surfaces as a continuum limit of random triangulations \cite{David:1984tx,Ambjorn:1985az,Kazakov:1985ds,Kazakov:1985ea,DiFrancesco:1993cyw}. 
This relationship between double-scaled matrix integrals  and 2d gravity has been studied extensively, and appears to provide a quantitative definition of the simplest universality classes of two-dimensional quantum gravity\cite{Brezin:1990rb,Douglas:1989ve,Gross:1989vs}.

This subject is currently the focus of renewed attention due to interest in the simplest model of 2d gravity, the Jackiw-Teitelboim model \cite{Jackiw:1984je,Teitelboim:1983ux}, and its relation to black holes \cite{Almheiri:2014cka, Engelsoy:2016xyb, Maldacena:2016upp, Jensen:2016pah} and the SYK model \cite{Sachdev:1992fk, Sachdev:2010um, KitaevTalks, Maldacena:2016hyu}. 
In \cite{ Saad:2019lba}, it was shown that the Euclidean path integral of this theory is equal to a particular double-scaled matrix integral, and in \cite{Maldacena:2019cbz,Cotler:2019nbi,Stanford:2019vob,Okuyama:2019xbv,Blommaert:2019wfy,Johnson:2019eik} some generalizations were considered.

The purpose of this paper is to show how to incorporate global symmetries in this correspondence between 2d gravity and matrix integrals. In order to do so, we include dynamical gauge fields in the bulk gravity theory. One motivation for this is as a warmup for discussing higher dimensional gravity theories (which would include dynamical gauge fields in a reduction to 2d), or gravity theories with higher supersymmetry. Another motivation is to explore the effect of 't Hooft anomalies in random matrix theory. Finally, we expect that aspects of this setup (although not the exact connection to random matrix theory) will be relevant for the bulk description of SYK-like theories with global symmetry \cite{Sachdev:2015efa, Gross:2016kjj, Fu:2016vas, Davison:2016ngz, Choudhury:2017tax, Moitra:2018jqs, Sachdev:2019bjn, Gu:2019jub}.

We will now summarize the paper.

In \hyperref[sec:RMT]{{\bf section two}}, we give a definition of random matrix theory including some global symmetry group $G$ that commutes with the random matrix (which we think of as a Hamiltonian). The Hilbert space breaks apart into different representations of $G$, and essentially, one can define separate random Hamiltonians acting in these subspaces. More precisely, we study correlation functions in the matrix ensemble of $\langle \tr e^{-\beta_1H}\dots \tr e^{-\beta_n H}\rangle_c$ where $H$ is the full Hamiltonian, including $G$ symmetry. Each of the partition functions can be decomposed into a sum of traces over subspaces corresponding to particular representations, $\tr_{\mathcal{H}_r}(e^{-\beta H})$, with correlations
\begin{equation}
    \langle \tr_{\mathcal{H}_{r_1}}(e^{-\beta_1 H})\dots \tr_{\mathcal{H}_{r_n}}(e^{-\beta_n H})\rangle_c = \delta_{r_1,\dots , r_n}\text{dim}(r_1)^n \langle \tr e^{-\beta_1H^{(r_1)}}\dots \tr e^{-\beta_n H^{(r_1)}}\rangle_c.\label{structureInINTRO}
\end{equation}
Here $H^{(r_i)}$ is the Hamiltonian acting on the different copies of the representation $r_i$. It is drawn from an ordinary random matrix ensemble without global symmetry. The expectation value on the RHS has a ``genus'' expansion, with coefficients determined by the loop equations \cite{migdal1983loop,eynard2004all} of random matrix theory.

In \hyperref[sec:BF]{{\bf section three}} we discuss the bulk gauge theory that will be added in order to reproduce the random matrix correlations of the form (\ref{structureInINTRO}). The partition functions of Yang-Mills and BF gauge theories on general Riemann surfaces were computed in \cite{Migdal:1975zg, Rusakov:1990rs, Fine:1990zz, Fine:1991ux, Blau:1991mp, Witten:1991we, Witten:1992xu}. We focus on the case of topological (BF-type) gauge theories, which are insensitive to local fluctuations of the geometry. This means that for each topology, the partition function of an arbitrary 2d gravity theory coupled to the gauge theory factorizes into the partition function of the gravity theory itself, times a partition function in the BF gauge theory. 

To compare to the random matrix observables (\ref{structureInINTRO}), we need to study the combined gravity theory on a space with $n$ boundaries, with particular boundary conditions related to the values of $\{\beta_i\}$ and the representations $\{r_i\}$. To match the results at a given order $g$ in the ``genus'' expansion of the matrix integral, we study the gravity theory on a space of genus $g$. As we show using \cite{Migdal:1975zg, Rusakov:1990rs, Fine:1990zz, Fine:1991ux, Blau:1991mp, Witten:1991we, Witten:1992xu}, the BF gauge theory partition function on this space with the appropriate boundary conditions is
\begin{equation}
    Z_{g}^{\text{gauge}}(\beta_1,r_1;\dots;\beta_n,r_n) = \delta_{r_1, \dots , r_n} \left(\text{dim}(r_1)\right)^n\left(\frac{\text{vol}(G)}{\text{dim}(r_1)}\right)^{2g+n-2}\prod_{j = 1}^n e^{-\beta_j c_2(r_1)/2}.\label{BFansINTRO}
\end{equation}
To get the full partition function, we multiply this by the partition function of the original 2d gravity theory without gauge fields.

\begin{figure}[t!]
    \centering
        \includegraphics[width=0.4 \textwidth]{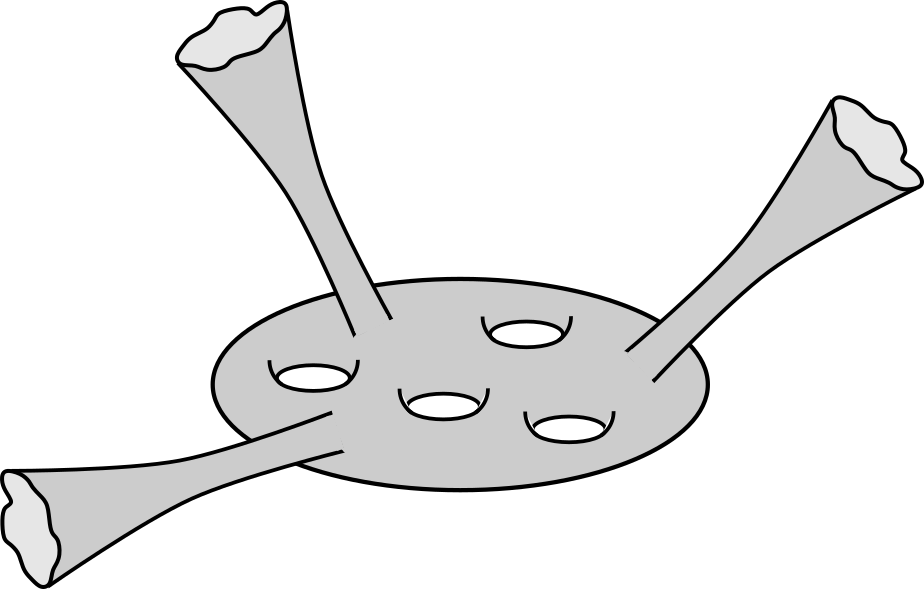}
    \caption{A cartoon of the geometry that computes $Z_{g}(\beta_1,r_1; \ldots; \beta_n,r_n)$. There is a genus $g$ surface glued to $n$ ``trumpets" associated to the insertions of $\tr_{\mathcal{H}_r}(e^{-\beta H})$ on the RMT side. The boundary conditions at the ends of the trumpets are determined by $\beta$ and $r$.}
    \label{fig:zgn}
\end{figure}

In \hyperref[sec:comparison]{{\bf section four}}, we show that (\ref{structureInINTRO}) and (\ref{BFansINTRO}) are compatible with each other, and that (\ref{BFansINTRO}) indicates a particular relationship between the random matrix theories that describe each subspace. In particular, the sector labelled by the irreducible representation $r$ has a density of states proportional to $(\dim(r))^2$. One factor of $\text{dim}(r)$ arises from the obvious spectral degeneracy due to the $G$ symmetry of the system. However,
the second factor of $\dim(r)$ implies that the ranks of the random matrices in the fixed representation sectors are proportional to $\text{dim}(r)$ as well. A possible explanation of the second factor of $\dim(r)$ is that the Hilbert space carries the action of a second copy of $G$ that commutes with the first $G$-symmetry but does not commute with the Hamiltonian.\footnote{
This is reminiscent of the construction in \cite{Lin:2019qwu} of gauge-invariant SL$_2$ operators that act on the Hilbert space of JT gravity coupled to matter, but do not commute with the Hamiltonian.}
We also find that the ground state energy in each sector is proportional to the quadratic Casimir $c_2(r)$. This result is clearly model dependent, but does reproduce the dependence of the ground state energy on the charge in the complex SYK model \cite{Gu:2019jub}.

In \hyperref[sec-anomalies]{{\bf section five}}, we discuss a 
generalization of this calculation, which immediately suggests itself in the gauge theory language but is rarely considered in the random matrix literature. This is to consider the effect of 't Hooft anomalies in the global symmetry $G$ on the random matrix ensemble. 
In the gauge theory formulation, this is achieved by including particular discrete topological terms in the bulk action. 
The resulting partition functions are still calculable, and can be expressed in terms of restricted classes of projective representations of the group $G$.
We find that the dual matrix integral again factors into distinct sectors, each now corresponding to a genuinely projective representation of $G$. 
These results dovetail nicely with recent discussions in the QFT literature \cite{Gaiotto:2017yup, Komargodski:2017dmc, Cordova:2019jnf} and are relevant for the complex SYK model, which exhibits a mixed anomaly between the global $U(1)$ symmetry and charge conjugation when the number of Dirac fermions is odd \cite{Fu:2016vas, Stanford:2017thb}.

In \hyperref[sec:timereversal]{{\bf section six}}, as a final variant, we also consider matrix ensembles with a $G$ symmetry and time-reversal invariance. 
In the bulk, this corresponds to summing over non-orientable surfaces. 
In the case without any 't Hooft anomalies, we find that the dual matrix integral is GOE or GSE like for self-conjugate representations of $G$, and contains degenerate pairs of GUE blocks for pairs $(r,\overline{r})$ with $r \neq \overline{r}$. 
We also study an example with a mixed anomaly between an internal $\mathbb{Z}_2$ symmetry and time reversal, and find two degenerate GUE-like sectors that are exchanged by the time reversal symmetry, even though both the irreps of $\mathbb{Z}_2$ are self-conjugate.
The interpretation is that, in the presence of the anomaly, the time-reversal operation and the $\mathbb{Z}_2$ operator \emph{anti}-commute, rather than commute.

In \hyperref[sec:discussion]{{\bf section seven}} we conclude, and in \hyperref[app:diskAndTrumpet]{{\bf appendix A}} we give a detailed computation of some BF theory path integrals that are used in the main text. In \hyperref[sec:Abelian]{{\bf appendix B}} we discuss the details of the $U(1)$ case as an illustrative example, and in \hyperref[sec:yangmills]{{\bf appendix C}} we discus the case where the bulk gravity theory is the JT model, where simpler computations with standard Yang-Mills theory are possible.

\noindent \emph{Note added:} After our analysis of the anomaly-free case was completed, we learned of the paper \cite{Iliesiu:2019lfc} which has partial overlap with our section \ref{sec:BF} and gives an interesting interpretation of the partition functions in terms of matrices whose entries are functions on the group. 
In the final stages of our project, we also learned of the paper \cite{Aminov:2019hwg} which has overlap with a special case of our results for the partition functions of $PSU(N)$ gauge theories with 't Hooft anomaly.

\section{Expectations from random matrix theory}\label{sec:RMT}

In the original application of random matrix theory to physics, one views the random matrix as a model for a quantum Hamiltonian $H$, and one defines an ensemble that is ``as random as possible'' given the symmetries of $H$. For example, without imposing any symmetries at all, a good notion of ``as random as possible'' is to integrate over all Hermitian matrices 
\begin{equation}
    \int \mathrm{d}H e^{-L \Tr V(H)} \, ,\label{intdef}
\end{equation}
where $\mathrm{d}H$ is a flat measure on the $L^2$ independent real components of an $L\times L$ Hermitian matrix, and $V(H)$ is a ``potential'' function that determines the leading distribution of the eigenvalues, generalizing Wigner's semicircle.

In random matrix theory, the most important symmetry is the antiunitary time-reversal symmetry. 
The presence or absence of this symmetry changes the random matrix ensemble in a fundamental way. 
For example, if the Hamiltonian commutes with a time reversal symmetry $\sf{T}$ satisfying $\sf{T}^2 = 1$, then one can choose a basis in which $\sf{T} = \sf{K}$, where $\sf{K}$ is simply the complex conjugation operator. 
$\sf{T}$-symmetry then implies that the components of $H$ are real, and the relevant notion of ``as random as possible'' replaces the integral $\mathrm{d}H$ over components of a Hermitian matrix with an integral $\mathrm{d}H$ over the independent components of a real symmetric matrix. 
This turns out to change the random matrix integral in rather important ways.

Ordinary unitary symmetries play a less fundamental role than time-reversal. 
The reason is that there is a somewhat trivial modification of ``as random as possible'' that incorporates unitary symmetries. 
To explain this, let $G$ be a group of unitary symmetries that commutes with $H$. 
Then the Hilbert space on which $H$ acts decomposes into a direct sum of subspaces corresponding to irreducible representations of $G$:
\begin{equation}\label{eq:Decomp}
    \mathcal{H} = \bigoplus_{r}\mathcal{H}_r.
\end{equation}
Each $\mathcal{H}_r$ consists of the states that transform under the symmetry $G$ in representation $r$. 
Such states can be further grouped into multiplets of dimension $\text{dim}(r)$ that transform among themselves under the action of $G$. 
So we can choose a basis for the states of the full Hilbert space as
\begin{equation}
    |r,a;i\rangle,\label{aindex}
\end{equation}
where $r$ labels the representation, $a\in \{1,\dots,\text{dim}(r)\}$ labels the state within a given multiplet, and $i$ labels the different multiplets. 
The $G$ symmetry of the Hamiltonian $H$ implies that
\begin{equation}
    \langle r,a;i|H|r',a';i'\rangle = \delta_{r,r'}\delta_{a,a'}H^{(r)}_{ii'}.\label{iindex}
\end{equation}
However, $G$ symmetry gives no constraints at all on the matrix $H^{(r)}$. 
One can therefore define a maximally random matrix ensemble with $G$ symmetry by taking the $H^{(r)}$ matrices for each representation $r$ to be independent random matrices.

To state this more precisely, we will have to introduce some notation, starting with the case with no $G$ symmetry. In this paper, we will focus on aspects of the random matrix $H$ that depend only on its eigenvalues $\lambda_1,\dots,\lambda_L$. A convenient observable of this type is the ``thermal partition function''
\begin{equation}
    Z(\beta,\{\lambda\}) = \Tr e^{-\beta H} =  \sum_{j = 1}^Le^{-\beta \lambda_j}.
\end{equation}
Below, we will leave the dependence on the eigenvalues $\{\lambda\}$ implicit and use the simpler notation $Z(\beta)$. 
We will also denote expectation values in the matrix integral (\ref{intdef}) using the notation $\langle \cdot \rangle$. 
An interesting set of observables are the connected correlation functions (or cumulants) of a product of $n$ partition functions. 
A basic fact in random matrix theory is that such observables have an asymptotic $1/L$ expansion of the form
\begin{equation}
    \langle Z(\beta_1)\dots Z(\beta_n)\rangle_c\simeq \sum_{g = 0,\frac{1}{2},1,\dots}^\infty \frac{Z_g(\beta_1\dots \beta_n)}{L^{2g+n-2}}.
\end{equation}
The summation variable $g$ is referred to as the ``genus'' and in general it takes both integer and half-integer values. 
In the simplest hermitian matrix integral, it takes only integer values. 
The coefficient functions $Z_g(\beta_1\dots \beta_n)$ give the $1/L$ expansion of the matrix integral, and they can be computed efficiently using a set of equations known as the ``loop equations.''\footnote{The loop equations were introduced in \cite{migdal1983loop}  and have been streamlined significantly over time, culminating in \cite{eynard2004all}. For a review, see \cite{eynard2015random} or section 4 of \cite{Stanford:2019vob}.}

There are two pieces of input data to the loop equations, and these two pieces of data specify what we mean by a matrix integral. 
The first piece is a discrete choice of symmetry class, in particular whether there is a time-reversal symmetry and if so whether $\sf{T}^2 = +1$ or $-1$. 
This piece of data determines what we mean by $\mathrm{d}H$ in (\ref{intdef}). 
The second piece of data is the potential $V(H)$.
In practice, it is more convenient to give an equivalent piece of data, which is the leading large $L$ approximation to the density of eigenvalues:
\begin{equation}
    \rho_0(x) = \lim_{L\rightarrow \infty}\frac{1}{L} \, \langle\sum_{i = 1}^L \delta(x-\lambda_i)\rangle.
\end{equation}
One can solve for $\rho_0(x)$ in terms of the potential and vice versa. 
In the simplest (``one cut'') matrix integrals, $\rho_0(x)$ is supported in a single interval of the real axis. 
Note that $\rho_0(x)$ is normalized so that its integral is one.  
The leading approximation to the total density of eigenvalues has an extra factor of $L$:
\begin{equation}
    \rho_0^{\text{total}}(x) = L\rho_0(x).
\end{equation}
Given these two pieces of data (the symmetry class and $\rho_0(x)$), a universal recursion relation that follows from the loop equations determines all of the $Z_g(\beta_1\dots \beta_n)$ coefficients \cite{eynard2004all}.

The types of matrix integrals that are related to 2d gravity are not quite of the form (\ref{intdef}). 
Instead, they are ``double-scaled'' matrix integrals. 
Formally, these are integrals in which the interval of support of the leading distribution of eigenvalues is the half-line $x\ge x_0$, and we relax the requirement of normalizability of $\rho_0(x)$. More precisely, they can be obtained as limits of ordinary matrix integrals, where we take $L\rightarrow \infty$ and adjust the potential $V(H)$ in such a way that pointwise 
\begin{equation}
    \rho_0^{\text{total}}(x) \rightarrow e^{S_0}\rho_0(x),
\end{equation}
where $\rho_0(x)$ is supported on the half-line $x\ge x_0$, and is not normalized. 
The parameter $e^{S_0}$ is the leftover parameter, which plays a role analogous to $L$ in ordinary matrix integrals. 
In particular, the double-scaling procedure commutes with the loop equations, but the expansion becomes a series in $e^{S_0}$ rather than $L$:
\begin{equation}
    \langle Z(\beta_1)\dots Z(\beta_n)\rangle_c\simeq \sum_{g = 0,\frac{1}{2},1,\dots}^\infty \frac{Z_g(\beta_1\dots \beta_n)}{(e^{S_0})^{2g+n-2}}.
\end{equation}
An important fact for this paper follows from the form of this expansion. 
If we take a double-scaled random matrix theory and rescale $\rho_0(x)\rightarrow \lambda \rho_0(x)$, then the expansion coefficients will transform as
\begin{equation}
    Z_g(\beta_1\dots\beta_n) \rightarrow \frac{1}{\lambda^{2g+n-2}}Z_g(\beta_1\dots \beta_n).\label{transform}
\end{equation}
To see this, note that both rescalings can be accomplished by shifting $S_0\rightarrow S_0 + \log\lambda$.

To state this more precisely, it is helpful to define a thermal partition function restricted to a given representation sector
\begin{equation}
    Z(\beta,r) := \tr_{\mathcal{H}_r}e^{-\beta H} = \text{dim}(r)\tr e^{-\beta H^{(r)}}.\label{dimrconversion}
\end{equation}
The trace in the final expression is over the $i$ index in (\ref{iindex}) that labels the different multiplets within the subspace $\mathcal{H}_r$. 
Independence of the $H^{(r)}$ matrices corresponding to different sectors implies a ``diagonal'' property in the representation label,
\begin{equation}
    \langle Z(\beta_1,r_1)\dots Z(\beta_n,r_n)\rangle_c = 
    \delta_{r_1, \dots , r_n}\text{dim}(r_1)^n \langle \tr e^{-\beta_1H^{(r_1)}}\dots \tr e^{-\beta_n H^{(r_1)}}\rangle_c.\label{structureIn}
\end{equation}
One can then define a random matrix ensemble in the presence of $G$ symmetry by saying that the expectation value on the RHS is an expectation value in an ordinary random matrix ensemble with no $G$ symmetry, as defined previously. 
In particular, associated to each sector $r$ will be a function $\rho_0^{(r)}(x)$ that characterizes the leading density of eigenvalues of the matrix $H^{(r)}$. 
The expectation value on the RHS will then have a genus expansion with coefficients determined by $\rho_0^{(r)}(x)$ and the loop equations.

In the next two sections, we will show how the structure in (\ref{structureIn}) arises in 2d gravity once we incorporate a bulk version of $G$ symmetry.

\section{Gauge theory partition functions}
\label{sec:BF}
We would like to start with a 2d gravity theory that is dual to (or at least approximated by) a random matrix ensemble without $G$ symmetry, and somehow modify it to make it dual to a random matrix ensemble with $G$ symmetry. 
For example, we can start with the correspondence between JT gravity and a particular double-scaled matrix integral. 
This can be viewed as a type of disorder-averaged AdS/CFT correspondence. 
In AdS/CFT duality generally, the presence of a global symmetry in the boundary theory means that the bulk theory should have a gauge symmetry. 
So, at least tentatively, in order to describe a random matrix ensemble with $G$ symmetry, we should include a dynamical $G$ gauge theory in the bulk.

We will choose the bulk action to be that of the simplest 2d gauge theory, which is the topological BF theory. 
Boundary terms and boundary conditions will be discussed in section \ref{sec-bdy-terms-conditions}.
The fields in this theory consist of a gauge field $A$ with field strength $F =\mathrm{d}A + A\wedge A$, and an adjoint-valued scalar field $B$. 
The action on a closed 2d manifold $\Sigma$ is
\begin{equation}
    I = -\mathrm{i} \int_\Sigma \tr(BF).
\label{eq:bfaction}
\end{equation}
The integral over the field $B$ is a Lagrange multiplier that imposes $F = 0$, 
so the path integral localizes to an integral over flat connections. In general, there is a moduli space of such connections, and the path integral computes the volume of this space, with respect to a measure that we will describe in more detail below.

\subsection{Path integral on a closed surface}
\label{sec-closedsurfaces}

The path integral can be computed efficiently by decomposing the surface into three-holed spheres, as explained in \cite{Witten:1991we}.\footnote{Other useful references treating relevant aspects of 2d Yang-Mills and BF theory include \cite{Migdal:1975zg, Rusakov:1990rs, Fine:1990zz, Fine:1991ux, Blau:1991mp, Witten:1992xu, Thompson:1992hv, Blau:1993hj, Blau:1995rs, Moore:1994dk, Cordes:1994fc}.}  
A three-holed sphere is a space with the topology of a sphere with three disks removed. 
To compute the path integral on such a space, we need boundary conditions for each of the $S^1$ boundaries. 
A convenient choice is to fix the holonomy $U = P\exp(-\int_\gamma A)$ around each circle. 
Under a gauge transformation that acts nontrivially at the boundary, the holonomy transforms by conjugation, so gauge invariance implies that the path integral with fixed holonomy will be a ``class function,'' meaning a function of $U$ that is conjugation-invariant. 
On the three-holed sphere, the path integral of the BF theory is a class function of the three holonomies
\begin{equation}\label{threeV}
  \Psi_0(U_1,U_2,U_3),
\end{equation}
where the subscript indicates genus zero. 

One can write a simple expression for (\ref{threeV}), as shown in \cite{Witten:1991we}.
The expression is particularly simple when expressed in the basis of characters. 
The characters $\chi_r(U) = \tr_r(U)$ for different representations provide a complete orthonormal basis of class functions:
\begin{equation}
    f(U) = \sum_{r}f(r) \chi_r(U), \hspace{20pt} f(r) = \int_G\frac{\mathrm{d}U}{\vol(G)}\overline{\chi_r(U)}f(U).
\label{projectiontor}
\end{equation}
The result of \cite{Witten:1991we} is that, in the representation basis, the three-holed sphere path integral is
\begin{equation}
   \Psi_0(r_1,r_2,r_3)=\frac{\text{vol}(G)}{\text{dim}(r_1)}\delta_{r_1,r_2,r_3}.\label{threeHoled}
\end{equation}
In particular, the result is ``diagonal.'' 

To compute the path integral of the BF theory on an arbitrary orientable closed surface, one simply builds the surface up by gluing together three-holed spheres.
Suppose that we have two spaces with $S^1$ boundaries that we want to glue together, and that the path integrals on the two spaces with fixed holonomy on the $S^1$ are  $\psi(U)$ and $\phi(U)$, respectively. Then the path integral on the connected space is
\begin{equation}
    \int \frac{\mathrm{d}U}{\vol(G)}\psi(U^{-1})\phi(U) = \sum_r \overline{\psi(r)}\phi(r).\label{gluingrule}
\end{equation}
The $U^{-1}$ in one factor is due to the fact that if we pick a consistent orientation of the 2d surface, and cut it along an $S^1$, then the opposite orientation of the $S^1$ will be induced on the two cut pieces.

To build a closed surface of genus $g>1$, one can glue together $2g-2$ three-holed spheres. Using (\ref{threeHoled}) and (\ref{gluingrule}), one finds that the genus $g$ partition function in the BF theory is
\begin{equation}\label{Zclosed}
Z_{g}^{\text{gauge}} = \sum_{r}\left(\frac{\text{vol}(G)}{\text{dim}(r)}\right)^{2g-2}.
\end{equation}

\subsection{Boundary conditions}
\label{sec-bdy-terms-conditions}
To make contact with random matrix theory, we would like to understand what gauge theory object is dual to an insertion in the matrix integral of
\begin{equation}
    \tr (e^{-\beta H} V) = \sum_r\frac{\chi_r(V)}{\dim(r)}Z(\beta,r).\label{Vinsert}
\end{equation}
The $Z(\beta,r)$ appearing here is a matrix integral partition function in a fixed representation sector, and was defined in (\ref{dimrconversion}). $V\in G$ is a group element, and we will sometimes write it as $V = e^{\beta \mu}$, with $\mu$ a chemical potential.\footnote{Here $\mu$ is a Lie algebra element, and is antihermitian in order for $e^{\beta \mu} \in G$. The standard hermitian chemical potential corresponds to a continuation where $e^{\beta \mu}$ is in the complexification of $G$.} 

In the bulk gauge theory, this insertion corresponds to the requirement that the 2d bulk space should have a boundary with boundary conditions determined by $\beta$ and $V$, in a way that we will describe below. In fact, to compare the gauge theory and matrix integral formulas, it will be convenient to directly work with insertions of $Z(\beta,r)$, with fixed representation $r$. These can be obtained by taking a weighted integral over $V$:
\begin{equation}
    Z(\beta,r) = \dim(r)\int \frac{\mathrm{d}V}{\vol(G)}\, \overline{\chi_r(V)}\, \tr( e^{-\beta H}V).\label{projectOnto}
\end{equation}
This insertion can be described as another type of boundary condition in the bulk gauge theory, characterized by $\beta$ and $r$.

We would like to compute the path integral of the gauge theory on arbitrary orientable surfaces, with boundaries of the type that correspond to insertions of $Z(\beta,r)$. 
This path integral can be constructed using a gluing procedure. 
One important component in the gluing is the three-holed sphere discussed above. We will also need the path integral on a ``trumpet'' geometry, with the $Z(\beta,r)$ boundary conditions at one end, and the gluing boundary condition at the other end. Finally, as a special case, we will also need the path integral on the disk with $Z(\beta,r)$ boundary conditions on the boundary. 
The disk and trumpet path integrals are computed from the BF theory perspective in appendix \ref{app:diskAndTrumpet}. We will discuss them from a related Yang-Mills perspective below. Either way, the results are
\begin{align}
    Z_D^{\text{gauge}}(\beta,r) &=  \frac{\text{dim}(r)^2}{\text{vol}(G)} e^{-\beta c_2(r)/2},\label{disk}\\
    Z_T^{\text{gauge}}(\beta,r;r') &= \delta_{r,r'}\,\text{dim}(r)e^{-\beta c_2(r)/2}.\label{trumpet}
\end{align}
For $Z_T^{\text{gauge}}$, the final $r'$ argument labels the representation at the gluing end of the trumpet.

The trumpet path integral, together with the formula for the three-holed sphere above, can be used to give the partition function of the BF theory with $Z(\beta,r)$ boundary conditions, and an arbitrary connected topology. 
The only fact that one needs to know is that a genus $g$ surface with $n$ boundaries can be decomposed into $2g+n-2$ three-holed spheres. 
Since the three-holed sphere is diagonal in representation (see eq.~(\ref{threeV})), and the trumpet is also diagonal in representation (see eq.~(\ref{trumpet})), all representations will have to be equal. 
Multiplying together the factors for $n$ trumpets and $2g+n-2$ three-holed spheres, one finds
\begin{equation}
    Z_{g}^{\text{gauge}}(\beta_1,r_1;\dots;\beta_n,r_n) = \delta_{r_1, \dots , r_n} \left(\text{dim}(r_1)\right)^n\left(\frac{\text{vol}(G)}{\text{dim}(r_1)}\right)^{2g+n-2}\prod_{j = 1}^n e^{-\beta_j c_2(r_1)/2}.\label{BFans}
\end{equation}
This formula is actually the only thing we will need to know about the bulk gauge theory in order to match to random matrix theory. Plugging in $g = 0$ and $n = 1$, one finds that this formula also gives the right answer for the special case of the disk (\ref{disk}).

The rest of this section should be regarded as optional. We will describe the boundary conditions that correspond to insertions of (\ref{Vinsert}). Because they are slightly unusual from the perspective of 2d gauge theory, we will discuss them from two different perspectives below. Detailed computations in the BF perspective are in appendix \ref{app:diskAndTrumpet}.

\subsubsection*{Yang-Mills theory}
It will be helpful to think about a generalization of the BF theory to a Yang-Mills theory with a position-dependent coupling constant $e^2(x)$:
\begin{align}
    I_{\text{YM}} = - \i \int \Tr({\bf BF}) - \frac{1}{2} \int\mathrm{d}^2x\sqrt{g} e^2(x)\Tr {\bf B^2}\, .
\label{eq-ymaction}
\end{align}
Here and below, we use boldface for the Yang-Mills fields, to distinguish them from the BF fields in the discussion below. After integrating out ${\bf B}$, one gets a standard Yang-Mills theory. A simple boundary condition is the ``gluing'' boundary condition discussed above, where we fix the holonomy $V$ of the gauge field around an $S^1$ boundary. To relate this to the dual boundary theory, we interpret $V$ as the holonomy of a background gauge field on the boundary circle, which is to say that this $V$ is the same as the $V$ in the insertion introduced in (\ref{Vinsert}).

The theory (\ref{eq-ymaction}) was solved in \cite{Witten:1991we} for an arbitrary $e^2(x)$, and the answer is simple to state. In the discussion of the BF theory, we saw that the three-holed sphere is ``diagonal'' in the representation basis, so the partition function can be written as a single sum over representations. The effect of the $e^2(x)$ term is simply to add an additional weighting 
\begin{equation}
    \exp\left(-\frac{c_2(r)}{2}\int \mathrm{d}^2 x \sqrt{g}e^2(x)\right).
\end{equation}
We will choose the function $e^2(x)$ to be concentrated in a narrow annular region near each true boundary (as opposed to artificial boundaries that are included in e.g.~the pants decomposition), so that this weighting factor is
\begin{equation}
    \exp\left(-\frac{c_2(r)}{2}\int \mathrm{d}^2 x \sqrt{g}e^2(x)\right) \to \exp\Big(-\frac{c_2(r)}{2}(\beta_1+\dots+\beta_n)\Big),
\end{equation}
where $\beta_j$ is the renormalized length of the $j$-th boundary. With this choice of $e^2(x)$, the disk and trumpet answers in (\ref{disk}) and (\ref{trumpet}) follow from results in \cite{Witten:1991we}.

The physics of this Yang-Mills setup is as follows. In the bulk of the space, we have $e^2(x) = 0$, so the $B$ field acts as a Lagrange multiplier that forces the gauge field to be flat. However, in a small region near the boundary, $e^2(x)$ is nonzero, and the gauge field can fluctuate. This fluctuation is the off-shell ``particle-on-a-group'' mode that has been identified in the low-energy limit of the SYK model with global symmetry \cite{Sachdev:2015efa, Gross:2016kjj, Fu:2016vas, Choudhury:2017tax, Moitra:2018jqs, Sachdev:2019bjn,Liu:2019niv, Mertens:2019tcm}. Because the gauge field is flat everywhere in the bulk of the space, one would expect that the somewhat unusual Yang-Mills theory we just described could be understood as a boundary condition in a conventional BF theory. We turn to this next.

\subsubsection*{BF theory with a non-topological boundary}
The same particle-on-a-group theory that we just discussed has also been derived from pure BF theory with a particular boundary condition in \cite{Blommaert:2018oro}. Generalizing this slightly to include a chemical potential, the theory is 
\begin{align}
    I= 
    -\i\int_{\Sigma}\tr BF -\frac{1}{2}\int_{\partial \Sigma} \mathrm{d}u\tr\Big[(A_u + \mu)^2\Big]\, ,
    \label{eq:fullaction}
\end{align}
together with the boundary condition
\begin{equation}
    B = \mathrm{i} (A_u + \mu)\big|_{\partial \Sigma}\, .
    \label{eq:boundarycondition}
\end{equation}
Here $u$ is a renormalized time coordinate along the boundary, and it runs from zero to $\beta$. The quantity $\mu$ is a chemical potential, as we will see in a moment. Note that although the boundary conditions relate $B$ to $A_u$, they do not determine the value of $B$ itself, and the fluctuations in $B$ become the particle-on-a-group mode \cite{Blommaert:2018oro}.

An important difference between these boundary conditions and the Yang-Mills ones is that (\ref{eq:boundarycondition}) is not gauge-invariant. So, in the BF description, we restrict the gauge transformations to act trivially at the boundary. The asymptotically constant would-be gauge transformations $A \rightarrow gAg^{-1} + g\mathrm{d}g^{-1}$  that do not vanish at the boundary become global symmetries. The corresponding Noether charge can be worked out from the action (\ref{eq:fullaction}):
\begin{equation}
    Q = A_u + \mu.
\end{equation}
From this it follows that for small variations $\mu \rightarrow \mu + \delta\mu$, the change in the action is
\begin{equation}
    \delta I = -\beta \tr(\delta\mu\;Q) \, ,\label{defchem}
\end{equation}
which means that $\mu$ is indeed the chemical potential.

In appendix \ref{app:diskAndTrumpet}, we work out the disk and trumpet formulas (\ref{disk}) and (\ref{trumpet}) in detail from the BF theory path integral with these boundary conditions. The fact that the answers agree with the Yang-Mills results from \cite{Witten:1991we} establishes in a roundabout way that the two descriptions are equivalent. We will also give a direct (although less precise) path-integral argument for their equivalence.

\subsubsection*{Relationship between the two descriptions}
In order to relate the two descriptions, it is convenient to consider the thin annulus near the boundary where the YM coupling is nonzero, and where the YM gauge field can fluctuate. 
Roughly, we identify the BF theory as living in the region inside the inner boundary of this strip, where the gauge field is forced to be flat. The fluctuating boundary mode of the BF theory will describe the YM gauge field in the remaining part of the geometry, which is the thin annulus itself.

In relating the theories, the following detail is important. Normally in YM theory, we fix the holonomy of the gauge field at the boundary in terms of the chemical potential $V = e^{\beta \mu}$, and we quotient by gauge transformations that are nonzero on the boundary. 
However, an equivalent prescription is to fix $A_\tau$ pointwise along the boundary, and {\it not} quotient by gauge transformations on the boundary. 
To relate the theories, we will take this perspective. Rather than defining the YM theory by $V = e^{\beta \mu}$, we will define it by
\begin{equation}{\bf A}_u|_{\partial\Sigma} =- \mu.\label{bwn}
\end{equation}

Now, the main idea is that the BF and YM gauge fields are the same in the interior of the geometry, and in the thin annular region, they differ by a term proportional to the boundary value of the $B$ field of the BF theory:
\begin{align}
    {\bf A} &= A + \mathrm{i}\frac{x}{\varepsilon} \, B \, \mathrm{d}u , \label{eq-aymbfrelation}\\
    {\bf F} &= \frac{\mathrm{i}}{\varepsilon}\, B\, \mathrm{d}x \wedge \mathrm{d}u + O(\varepsilon^0).
\end{align}
Here $x$ is a coordinate that goes from $x=0$ (inner boundary of the thin annulus) to $x=\varepsilon$ (outer boundary of the thin annulus). The motivation for this term linear in $B$ is that it makes the BF boundary condition (\ref{eq:boundarycondition}) consistent with the YM boundary condition (\ref{bwn}). Substituting the above into the YM action, we find that in the thin annulus,
\begin{align}
   I=  -\int_{\text{annulus}} \tr\left( \i {\bf BF} + \frac{1}{2} \sqrt{g}e^2(x) \, {\bf B}^2\mathrm{d}^2x \right) = 
    -\int \mathrm{d}u\, \tr\left( - {\bf B} B + \frac{1}{2} {\bf B}^2 \right).
\end{align}
In going to the final expression, we integrated over the radial $x$ direction in the thin annulus, assuming that ${\bf B}$ is approximately constant in this small interval. After doing the Gaussian integral over ${\bf B}$, this becomes
\begin{equation}
  I \supset \int \mathrm{d}u \, \frac{1}{2} \tr(B^2) = -\frac{1}{2}\int \mathrm{d}u\tr\left[(A_u + \mu)^2\right],
\end{equation}
which is the action for the boundary mode in the BF description. So we see that after integrating out the ${\bf B}$ field of the YM description, the fluctuations of ${\bf A}$ in the thin annulus become the fluctuations in the boundary mode of the BF description.

\section{Comparison to random matrix theory}
\label{sec:comparison}

Now, let us try to compare the bulk gauge-theory result (\ref{BFans}) to random matrix theory.  As a starting point, we should remember that the BF or YM gauge theory is only part of the bulk theory. In addition, we have whatever ``seed'' 2d gravity theory is dual to the matrix integral without global symmetry. For example, this could be the $(2,p)$ minimal string, or JT gravity. Then the full partition function on genus $g$ with $n$ boundaries characterized by $\{\beta_i,r_i\}$ is 
\begin{equation}
    Z_g(\beta_1,r_1;\dots;\beta_n,r_n) = Z_{g}^\text{grav}(\beta_1,\dots, \beta_n)Z_{g}^{\text{gauge}}(\beta_1,r_1;\dots;\beta_n,r_n).\label{checked}
\end{equation}
Here $Z_{g}^{\text{grav}}(\beta_1\dots \beta_n)$ is the partition function of the 2d gravity theory without the bulk gauge theory included, and $Z_g^{\text{gauge}}$ is given in (\ref{BFans}). Note that this simple factorization is a result of the fact that the BF theory is topological in the bulk, and doesn't depend on the details of the metric or the bulk gravity theory.

We define a seed $\rho_0^{\text{grav}}(x)$ as the leading density of eigenvalues associated to $Z_{0}^{\text{grav}}$:
\begin{equation}
   Z_{0}^{\text{grav}}(\beta) = \int \mathrm{d}x\, e^{-\beta x}\rho_0^{\text{grav}}(x).\label{gravAns}
\end{equation}
Given this function, what is the leading density of eigenvalues of the full theory in a given representation sector? 
Due to the exact $\dim(r)$ degeneracy in sector $r$, we would like to write
\begin{equation}
    Z_0(\beta,r) =\dim(r) \int \mathrm{d}x\, e^{-\beta x}\rho_0^{(r)}(x)\, ,\label{ahj}
\end{equation}
with $\rho_0^{(r)}(x)$ to be interpreted as the leading density of distinct eigenvalues in the sector of the Hilbert space transforming in representation $r$, not including the degeneracy. In other words, $\rho_0^{(r)}(x)$ is the leading density of eigenvalues of $H^{(r)}$. Our formula for the disk (\ref{disk}) implies that
\begin{equation}
    Z_0(\beta,r) = Z_0^{\text{grav}}(\beta)\frac{\dim(r)^2}{\vol(G)}e^{-\beta c_2(r)/2}.
\end{equation}
Compatibility with (\ref{gravAns}) and (\ref{ahj}) then requires that
\begin{equation}
    \rho_0^{(r)}(x) = \frac{\text{dim}(r)}{\text{vol}(G)}\, \rho_0^{\text{grav}}(x-c_2(r)/2).
    \label{eq:rho0rx}
\end{equation}
So the leading density of eigenvalues in the different sectors are all described by the same function $\rho_0^{\text{grav}}(x)$, but with shifted ground state energies $c_2(r)/2$, and densities (of distinct eigenvalues) rescaled by a factor proportional to $\text{dim}(r)$. Note that this factor in the density of {\it distinct} eigenvalues combines with the exact $\text{dim}(r)$ degeneracy to imply that the {\it total} number of eigenvalues in each sector is proportional to $\text{dim}(r)^2$.

In random matrix theory, the function $\rho_0^{(r)}(x)$ completely determines the genus expansion of correlation functions of $\tr e^{-\beta H^{(r)}}$. And, because of the simple relationship between $\rho_0^{(r)}$ and $\rho_0^{\text{grav}}$, the prediction can actually be written in a simple way in terms of the correlation functions of the seed theory $Z_g^{\text{grav}}(\beta_1\dots \beta_n)$. The relationship is
\begin{equation}
    \langle \tr e^{-\beta_1 H^{(r)}}\dots \tr e^{-\beta_n H^{(r)}}\rangle\Big|_g =Z_g^{\text{grav}}(\beta_1\dots\beta_n) \left(\frac{\text{vol}(G)}{\text{dim}(r)}\right)^{2g+n-2} \prod_{j = 1}^ne^{-\beta_j c_2(r)/2}.
\end{equation}
Both of the multiplicative factors on the RHS have simple explanations. The factors of $e^{-\beta_j c_2(r)/2}$ account for the shift in the ground state energy by $c_2(r)/2$. The other factor involving vol$(G)/\dim(r)$ accounts for the rescaling of $\rho_0^{(r)}$ relative to $\rho_0^{\text{grav}}$, as explained in (\ref{transform}). 

Finally, using (\ref{structureIn}) to convert from the partition function of distinct eigenvalues $\tr e^{-\beta H^{(r)}}$ to the total partition function in each sector $Z(\beta,r)$, we find
\begin{align}
    Z_g(\beta_1,r_1;\dots;\beta_n,r_n)&=\langle Z(\beta_1,r_1)\dots Z(\beta,r_n)\rangle\big|_g\\
    &=\delta_{r_1, \dots , r_n}\dim(r_1)^n  \langle \tr e^{-\beta_1 H^{(r_1)}}\dots \tr e^{-\beta_n H^{(r_1)}}\rangle\big|_g\\ 
    &=\delta_{r_1, \dots , r_n}
    \dim(r_1)^nZ_g^{\text{grav}}(\beta_1\dots\beta_n) \left(\frac{\text{vol}(G)}{\text{dim}(r_1)}\right)^{2g+n-2} \prod_{j = 1}^ne^{-\beta_j c_2(r_1)/2}.\label{matrixAgreement}
\end{align}
This equation is precisely consistent with (\ref{checked}) and the answer for the BF theory in (\ref{BFans}). This establishes that if the seed 2d gravity theory was dual to a random matrix ensemble, then after coupling to BF theory, the result will be dual to a random matrix ensemble with $G$ symmetry.

\section{Global symmetries with anomalies}
\label{sec-anomalies}

In Section \ref{sec:RMT} it was assumed that, for a quantum mechanical ensemble with $G$ symmetry, the Hilbert space on which the Hermitian matrix $H$ acts decomposes into a direct sum of subspaces corresponding to the irreducible representations of $G$. However, since states in quantum mechanics are identified projectively, the total Hilbert space is only required to furnish projective representations of $G$. In principle, the decomposition of the Hilbert space in \eqref{eq:Decomp} could therefore include a sum over projective representations. As we will review, replacing the sum over ordinary representations by a particular class of these projective representations can be described by saying that the theory has an 't Hooft anomaly in the realization of the $G$ symmetry. 

The presence of this anomaly is a type of qualitative distinction in the realization of the $G$ symmetry. In the AdS/CFT context (and in the study of SPT phases), an 't Hooft anomaly in the boundary description can be described in the bulk by adding discrete topological terms that assign phase weighting to nontrivial gauge bundles. In this section, we will see that after including such terms, the bulk theory remains dual to a random matrix integral, but with a particular set of projective representations appearing, indicating the 't Hooft anomaly.

We will work with the Yang-Mills description of the theory, since it is straightforward to incorporate holonomies that do not belong to the connected component of the identity in $G$. The coupling will still be delta-function localized on the boundary, so the theory is topological away from the boundary.

\subsection{Non-trivial bundles and discrete $\theta$ terms}
\label{secbundleintro}

The partition function of a gauge theory with gauge group $G$ on a manifold $\Sigma$ involves an integral over gauge equivalence classes of connections on all principal $G$-bundles with base $\Sigma$. 
For topologically trivial $\Sigma$, the only class of $G$-bundle is $G\times \Sigma$, but for general $\Sigma$ and certain choices of $G$, the partition function separates into integrals over distinct topological classes of bundles. 
The definition of the quantum field theory requires an unambiguous rule for weighting these different sectors in a manner compatible with locality and unitarity. 
In previous sections, we have implicitly weighted all classes of bundles uniformly. When $G$ is a simply-connected Lie group, there are no non-trivial bundles and the point is moot. However, for generic $G$ and $\Sigma$ there exist distinct theories that differ from one another through the inclusion of discrete topological terms in the gauge theory action.

The allowed weighting factors for topological classes of $G$-bundles are classifiable \cite{Dijkgraaf:1989pz,Chen:2011pg} in any dimension, but for our purposes a more pedestrian approach is possible. 
Since our primary motivation is to introduce a sum over projective representations in the dual matrix ensemble, we will restrict attention to topological weighting factors that measure the obstruction to lifting a $G$-bundle to a $\widetilde{G}$-bundle, where $\widetilde{G}$ is a central extension of $G$.
Rather than providing an exhaustive treatment, we will specialize the general discussion to an interesting subclass of theories, and then illustrate the construction with a small number of representative examples.

\subsubsection*{Projective representations and central extensions}
A unitary \emph{projective representation} of a group $G$ on a vector space $\mathcal{H}$ is a homomorphism from $G$ to $PU(\mathcal{H})$.\footnote{See \S 53 of \cite{Curtis} or \S 20 of \cite{huppert} for a readable account of the necessary results on projective representations and central extensions.}
Given such a representation, one can pick a (non-unique) lift to $U(\mathcal{H})$, in which case the group law is obeyed up to a system of phases:
\begin{equation}
    U(g_1g_2)=\omega(g_1,g_2) U(g_1)U(g_2)\, , \qquad 
    \omega(g_1,g_2)\in U(1)\, ,  \qquad
    U(g)\in U(\mathcal{H}).
\end{equation}
The system of phases $\omega:G\times G \to U(1)$ defines a 2-cocycle in the group cohomology of $G$.
Since the lift is non-unique, each element $U(g)$ is only defined up to a $g$-dependent phase $U'(g)=f(g)U(g)$. Changing the lift sends
\begin{equation}
    \omega(g_1,g_2)\to \omega(g_1,g_2)f(g_1g_2)f^{-1}(g_1)f^{-1}(g_2),
\end{equation}
and quotienting by this redundancy defines a class $[\omega]\in H^2(G,U(1))$. If this class is trivial, the representation can be ``de-projectivized," or lifted to an honest linear representation of $G$, while a nontrivial class $[\omega]$ presents an obstruction to de-projectivization.

The \emph{central extension} of a group $G$ by an abelian group $C$ is defined by a short exact sequence
\begin{equation}
    1\to C \xrightarrow{i} \widetilde{G}\xrightarrow{\pi} G \to 1,
\end{equation}
where $C$ is a subgroup of the center of $\widetilde{G}$, $i$ is the inclusion map and $\pi: \widetilde{G} \to G$ is a surjective homomorphism. Central extensions of $G$ by $C$ can also be classified, up to isomorphism, by group cohomology. Each such extension corresponds to a class $[E]\in H^2(G,C)$. Therefore, isomorphism classes of central extensions of $G$ by $U(1)$ are in bijective correspondence with isomorphism classes of projective representations. 

In the case where $C$ is a proper subgroup of $U(1)$, there are generally more group extensions of $G$ by $C$ than there are distinct classes of projective representations with phase systems $\omega$ contained in $C$. 
However, there is always at least one group $\widetilde{G}$ whose linear representations contain all of the projective representations of $G$: the Schur covering group $\widetilde{G}_S$.\footnote{See, for example, \url{https://groupprops.subwiki.org/wiki/Schur_covering_group}, or \S 20 of the book \cite{huppert}.} 
The Schur covering group is a stem\footnote{A central extension $1 \to C \to \widetilde{G}\to G \to 1 $ is called a stem extension if $C$ is also contained in the commutator subgroup of $\widetilde{G}$.} extension of $G$ by the degree two group homology of $G$:
\begin{equation}
    1 \to H_2(G,\mathbb{Z}) \xrightarrow{i} \widetilde{G}_S 
    \xrightarrow{\pi} G\to 1\, .
\end{equation}
For a finite group, $H_2(G,\mathbb{Z})$ is the same as $H^2(G, U(1))$, and is also known as the \emph{Schur multiplier} of $G$.
The Schur covering group is not unique unless the group $G$ is perfect (equal to its commutator subgroup), but all Schur covering groups for a finite group $G$ are isoclinic.\footnote{See, for example, \url{https://groupprops.subwiki.org/wiki/Isoclinic_groups}.}
In particular, this implies that if $\widetilde{G}_1$ and $\widetilde{G}_2$ are two Schur covering groups of a finite group $G$, 
then $\widetilde{G}_1$ and $\widetilde{G}_2$ have the same number of representations of each dimension. 
In other words, the partition functions on closed surfaces of two 2D gauge theories whose gauge groups are isoclinic will be equal to each other.

When $G$ is a compact, connected semi-simple Lie group, the analog of the Schur covering group is the universal cover $\widetilde{G}$ of $G$. In this case, $\widetilde{G}$ is simply connected, and the quotient $\pi: \widetilde{G}\to G$ by a subgroup $\Gamma = \ker(\pi)$ of the center of $\widetilde{G}$ realizes $\widetilde{G}$ as a central extension of $G$ by $\Gamma$. All projective representations of $G$ can be realized as linear representations of $\widetilde{G}$, and the non-trivial classes $[\omega]$ correspond to representations of $\widetilde{G}$ for which some element of $\Gamma$ is represented non-trivially. 

Below, we will focus on the case with $\ker(\pi) \cong \mathbb{Z}_N$, with group elements $1,z,z^2,\dots,z^{N-1}$. Then in a given representation $\widetilde{r}$ of $\widetilde{G}$, the generator $z$ is equal to $e^{2\pi \mathrm{i}k'/N}$ for some $k'$ that depends on the representation $\widetilde{r}$. So in general
\begin{equation}
    \chi_{\widetilde{r}}(\widetilde{g}z) = e^{2\pi\mathrm{i}k'(\widetilde{r})/N}\chi_{\widetilde{r}}(\widetilde{g}).\label{nality} 
\end{equation}
We will follow a convention and refer to the value of $k'$ as the ``$N$-ality'' of the representation $\widetilde{r}$.\footnote{If $p$ is coprime with $N$, then $z^p$ also generates $\ker(\pi)$. Our choice of generator $z$ defines what we mean by $N$-ality of a representation of $G$, as well as the $G$-bundle with minimal topological number in the discussion that follows.}

\subsubsection*{Topological actions and partition functions for finite gauge groups}
For a finite gauge group $G$, each bundle on a closed Riemann surface $\Sigma$ of genus $g$ is completely specified by the holonomies around the non-contractible cycles in $\Sigma$, modulo overall conjugation by $G$. Each bundle therefore represents an element of $ \text{Hom}(\pi_1(\Sigma),G)/G$ and a unique topological class. 
Since there is a unique flat connection in each class, the possible partition functions of the theory are completely determined by the relative weights assigned to each topological class of bundle: 
\begin{equation}\label{topZ}
    Z_\Sigma = \frac{1}{|G|}\sum_{\gamma \in \text{Hom}(\pi_1(\Sigma),G)} e^{2\pi \i S(\Sigma,\gamma)}\, .
\end{equation}

The simplest choice of topological action, $S(\Sigma,\gamma)=0$, weights all bundles equally and simply counts the number of $G$-bundles on $\Sigma$, modulo conjugation. 
In general, we can allow $S(\Sigma,\gamma)$ to be multivalued, provided that the phase in \eqref{topZ} is well-defined. 
The possible choices of $S(\Sigma, \gamma)$ consistent with factorization can be classified using group cohomology \cite{Dijkgraaf:1989pz}. 
Each action corresponds to a class $[S]\in H^2(BG,U(1))$, where $BG$ is the classifying space for principal $G$-bundles.\footnote{$BG$ is the base space of the so-called ``universal bundle" $EG$. The topological classes of $G$ bundles over a manifold $M$ are classified by the homotopy classes of maps $M\xrightarrow{\rho} BG$, which can be used to pull back the bundle over $BG$ to a bundle over $M$. See section 2 of \cite{Dijkgraaf:1989pz} for a discussion aimed at physicists.} This cohomology ring is isomorphic to the group cohomology used in the classification of projective representations, $H^*(BG,U(1))=H^*(G,U(1))$, so that a choice of topological action also singles out a particular class of non-trivial projective representations of $G$. As we will see, the partition function of this theory in the representation basis can be expressed solely in terms of this particular class of projective representations.
Concretely, the classifying map  $\rho:\Sigma\to BG$ allows one to pull back cohomology $\rho_*:H^2(BG,U(1))\to H^2(\Sigma,U(1))$. Pairing the image of $[S]$ under this map with the fundamental homology cycle $[\Sigma]$ yields a topological action 
\begin{equation}\label{topAct}
    S(\Sigma,\gamma)=\langle \rho_*[S], [\Sigma]\rangle
\end{equation}
which depends only on the topological class of $G$ bundle (provided that $\Sigma$ has no boundary). 

We would like a practical way to evaluate the action \eqref{topAct} in concrete examples.
We will focus on finite groups $G$ such that $H^2(BG,U(1))$ equals $\mathbb{Z}_N$.
In this case, the group $G$ has $N$ distinct classes of projective representations. We choose a Schur covering group\footnote{The choice does not affect the answer since all Schur covering groups are isoclinic.} (in this case an extension of $G$ by $\mathbb{Z}_N$), and distinguish (and weight) $G$ bundles according to whether or not they lift to $\widetilde{G}$ bundles. 
For discrete $G$, each bundle is topological, and is therefore completely specified by its transition functions $g_{ij}$ on an open cover $\{ U_i\}$ of $\Sigma$.
On each triple intersection of the open cover, the transition functions of the $G$ bundle satisfy the cocycle condition
\begin{equation}
    g_{ij}g_{jk}g_{ki}=1.
\end{equation}
Picking a lift $G\to \widetilde{G}$, each transition function lifts to an element in $\widetilde{G}$: $g_{ij}\to \widetilde{g}_{ij}$. 
On a generic triple intersection, one will have
\begin{equation}
    \widetilde{g}_{ij}\widetilde{g}_{jk}\widetilde{g}_{ki}=\widetilde{c}_{ijk} \in \mathbb{Z}_N \, .
\end{equation}

A nonzero element of $\mathbb{Z}_N$ appearing in the cocycle condition represents a possible obstruction to lifting the $G$ bundle to a $\widetilde{G}$ bundle. 
Changing the lift of the transition functions can change the local values of the $\widetilde{c}_{ijk}$, but if the transition functions cannot be lifted in such a way so as to remove this obstruction globally, then the $G$-bundle does not lift to a $\widetilde{G}$ bundle and we expect the topological action \eqref{topAct} to be nonzero. Concretely, the assignment of an element of $\mathbb{Z}_N$ to each triple intersection (modulo local gauge transformations and changes of lift) defines a \v{C}ech 2-cocycle $[\omega_N]\in H^2(\Sigma,\mathbb{Z}_N)$ which can then be paired with $[\Sigma]$ to define an integer $\omega=\langle [\omega_N],[\Sigma]\rangle\in \mathbb{Z}_N$. 

Since we are assuming that $H^2(BG,U(1)) = \mathbb{Z}_N$, our theories are labelled by an integer $k \in \{0, \ldots, N-1\}$.\footnote{We note that, in general, the set of possible topological terms is labelled by elements of $(\ker \pi)^\star$, the dual group of $\ker \pi$, which is the group of homomorphisms from $\ker \pi$ into the circle group $U(1)$.}
To determine the effect of the topological term \eqref{topAct} on the partition function, we begin by grouping the $G$-bundles on $\Sigma$ according to the value of $\omega$:
\begin{equation}
    Z_\Sigma^{(k)} =\frac{1}{|G|}\sum_{\gamma} e^{2\pi \i k\omega(\gamma)/N}=\sum_{\omega=0}^{N-1}e^{2\pi \i k\omega/N} Z_{\Sigma,\,\omega}\, ,
\label{eq:aksigma2}
\end{equation}
where $\gamma \in \text{Hom}(\pi_1(\Sigma),G)$ and $Z_{\Sigma,\,\omega}$ denotes the partition function restricted to $G$-bundles on $\Sigma$ with topological number $\omega$. 
In order to determine $Z_{\Sigma  , \, \omega}$, it is convenient to introduce the partition function of the gauge theory with gauge group $\widetilde{G}$ with no topological weighting factor, which we denote $\widetilde{Z}_\Sigma$:
\begin{equation}
    \widetilde{Z}_\Sigma = \frac{1}{|\widetilde{G}|}\sum_{\gamma\in \text{Hom}(\pi_1(\Sigma),\widetilde{G})} 1 
    = |\widetilde{G}|^{2g-2} \sum_{\widetilde{r}} \, (\dim \widetilde{r})^{2-2g}\, .
\end{equation}
The final expression is the analog of (\ref{Zclosed}) for finite groups, and it is derived the same way \cite{Witten:1991we}. The irreducible representations of $\widetilde{G}$ are denoted by $\widetilde{r}$. Because $\widetilde{G}$ is a Schur covering group, all representations of $G$, projective or not, occur in the sum.

All $G$ bundles that do not lift to $\widetilde{G}$ bundles can be obtained from honest $\widetilde{G}$ bundles using a simple trick that involves excising a small disk $D$ from $\Sigma$ \cite{Witten:1991we,Witten:2000nv}. 
As we have seen, specifying a $G$ bundle for a finite group $G$ involves specifying holonomies $U_1, V_1, \ldots, U_g, V_g$ such that $(U_1 V_1 U_1^{-1} V_1^{-1})$ $\ldots$ $(U_g V_g U_g^{-1} V_g^{-1})$ $= 1_G$.
Now, we pick lifts $\widetilde{U}_i \in \pi^{-1} (U_i)$ and $\widetilde{V}_i \in \pi^{-1}(V_i)$ and consider the element 
\begin{align}
u = 
( \widetilde{U}_1 \widetilde{V}_1 \widetilde{U}_1^{-1} \widetilde{V}_1^{-1} ) \ldots 
( \widetilde{U}_g \widetilde{V}_g \widetilde{U}_g^{-1} \widetilde{V}_g^{-1} ) 
\in \widetilde{G}.
\end{align}
Since we began with an honest $G$-bundle, it is guaranteed that $u \in \ker \pi \simeq \mathbb{Z}_N$.
However, $u$ may not be equal to $1_{\widetilde{G}}$, in which case the bundle does not lift. The value of $\omega \in \mathbb{Z}_N$ in (\ref{eq:aksigma2}) is given directly by $u$.

Note that the $\widetilde{U}_i$'s and the $\widetilde{V_i}$'s become admissible holonomies for a $\widetilde{G}$ bundle if we cut out a small disk $D$ from $\Sigma$\footnote{Concretely, imagine the representation of $\Sigma$ as a $4g$-gon with the standard identification of the edges, and imagine cutting out the disk from the center of this polygon.} and impose a holonomy around $\partial D$ equal to $u^{-1}$. Therefore, an equivalent construction of  $G$-bundles with obstruction number $\omega$ simply starts with the collection of admissable  $\widetilde{G}$-bundles, excises a small disk in each, and imposes the holonomy $u^{-1}$ determined by $\omega$ around the excision. Since this action is performed on each $\widetilde{G}$ bundle, we can use the formula for the partition function of the $\widetilde{G}$-theory on a Riemann surface with one boundary \cite{Witten:1991we}  to  obtain
\begin{align}
    Z_{\Sigma, \, \omega} &=N^{1-2g}\widetilde{Z}_{\Sigma}(u)= 
   \frac{1}{N}|G|^{2g-2} \sum_{\widetilde{r}}
    (\dim \widetilde{r})^{2-2g}\, 
    \frac{\chi_{\widetilde{r}}(u^{-1})}{\dim \widetilde{r}} \, . \quad  \label{eq:zomegasigma}
    \end{align}
This formula requires some explanation. First, $u$ is the element of $\text{ker}(\pi)\simeq \mathbb{Z}_N$ that is given by $z^\omega$, where $z$ is the elementary generator of $\mathbb{Z}_N$. From (\ref{nality}), we therefore have
    \begin{align}
    \chi_{\widetilde{r}}(u^{-1}) &= 
    \dim (\widetilde{r})\, e^{-2\pi \i k'(\widetilde{r})\omega/N}\, ,
\end{align}
where $k'$ is the $N$-ality, discussed near (\ref{nality}). Finally, the factor $N^{1-2g} = (\vert \widetilde{G} \vert / \vert G \vert)^{1-2g}$ can be interpreted as the inverse of a $\mathbb{Z}_N$ partition function, which corrects for the fact that the $\widetilde{G}$ gauge theory overcounts configurations from the perspective of a $G$ gauge theory. In particular, each of the $2g$ holonomies can be given $N$ different lifts to $\widetilde{G}$.

Substituting (\ref{eq:zomegasigma}) into (\ref{eq:aksigma2}) and using the orthogonality relation
\begin{equation}
    \sum_{\omega=0}^{N-1}e^{2\pi \i \omega(k-k')/N} = N\delta_{k,k'} \, ,
\end{equation}
with $k,k' \in \{0, \ldots N-1\}$,
we conclude that $Z_\Sigma^{(k)}$ is a sum over only those representations of $\widetilde{G}$ with ``$N$-ality" $k$. Note that for $k = 0$, the only representations that contribute are linear representations of $G$, and the normalization factors are such that we recover the original $Z_\Sigma$.

\subsubsection*{Connected Lie Groups}
When $G$ is a connected, simply-connected Lie group, all principal $G$-bundles are topologically trivial and homeomorphic to $G\times \Sigma$. 
When $G$ is connected, but not simply-connected, non-trivial bundles exist and are in one-to-one correspondence with the elements of $\pi_1(G)$. This correspondence can be explained as follows. 
The universal cover $\widetilde{G}$ is a central extension of $G$ by $\pi_1(G)$. 
Because $\widetilde{G}$ is simply connected, there is only one class of bundle $\widetilde{E}=\widetilde{G}\times \Sigma$, and there is a simple procedure to construct the non-trivial $G$ bundles from $\widetilde{E}$ \cite{Witten:1991we,Witten:2000nv}, just as in the discrete group case.\footnote{See also \cite{Grignani:1997yg, Fine:1995it}.} 
Given the trivial bundle $\widetilde{E}$, one cuts out a small disk $D$ from $\Sigma$, and glues it back with a transition function $u\in \pi_1(G)$. 
This procedure defines a $G$ bundle $E_u$ which does not lift to a $\widetilde{G}$ bundle. 
To see this, note that because $\pi_0(G)=0$, the bundle is trivial when restricted to $D$ and to $\Sigma-D$. 
On each patch, $E_u$ therefore lifts to a $\widetilde{G}$-bundle. 
However, the overlap of these two patches is a circle, and the transition function $u:S^1\to G$ defines a closed path in $G$ but not in $\widetilde{G}$. 
The lifted transition functions are therefore not single-valued in $\widetilde{G}$ and $E_u$ is not homeomorphic to $\widetilde{E}$.

One can alternately diagnose this obstruction using a characteristic class. Given a $G$-bundle $E_u$, the transition functions on the open cover satisfy the cocycle condition on triple intersections
\begin{equation}
    g_{ij}g_{jk}g_{ki}=1 \,.
\end{equation}
$E_u$ lifts to a $\widetilde{G}$-bundle on each open patch, but the lifts of the transition functions will generically satisfy
\begin{equation}
       \widetilde{g}_{ij}\widetilde{g}_{jk}\widetilde{g}_{ki}=\widetilde{c}_{ijk}\in \pi_1(G)\, .
\end{equation}
Non-trivial $\widetilde{c}_{ijk}$ presents a potential obstruction to lifting $E_u$ to a $\widetilde{G}$-bundle. Changing the patchwise lift changes the local values of the $\widetilde{c}_{ijk}$, but if they cannot all be trivialized globally then the bundle does not lift. In fact, since the obstruction is topological, it can always be localized to a small patch as in the previous discussion. The assignment of an element of $\pi_1(G)$ to each triple intersection (modulo local gauge transformations and changes of lift) in the open cover of $E_u$ defines a \v{C}ech 2-cocycle $[\omega]\in H^2(\Sigma,\pi_1(G))$ which can be paired with $[\Sigma]$ to produce the topological action
\begin{equation}
    S=\langle [\omega],[\Sigma]\rangle\, ,
\end{equation}
which depends only on the topological class of the bundle $E_u$.

When $\pi_1(G) = \mathbb{Z}_N$, the analysis is the same as for the finite group case above. One finds
\begin{equation}
    Z_{\Sigma,\omega}= \frac{1}{N} (\text{vol}(G))^{2g-2} \sum_{\widetilde{r}}\text{dim}(\widetilde{r})^{2-2g} e^{-2\pi \i \omega k'(\widetilde{r})/N}\, .
\label{eq:zsigmausun}
\end{equation}
Summing over $\omega$ as in (\ref{eq:aksigma2}), one finds again that the $Z_{\Sigma}^{(k)}$ is a sum over representations of $\widetilde{G}$ with $N$-ality equal to $k$.

\subsection{Manifolds with boundary and the anomaly}
\label{sec:anombound}

The characteristic classes that define the topological terms are well-defined and gauge invariant on closed manifolds: they depend only on the bundle, rather than a particular connection on the bundle. In particular,
given some representative $\omega_N$ of the class $[\omega_N]\in H^2(\Sigma,\mathbb{Z}_N)$, a variation of the connection or a gauge transformation induces a shift by a coboundary
\begin{equation}
    \omega_N\to \omega_N+\delta b
\end{equation}
 which integrates to zero on a closed surface: \begin{equation}
   \delta \omega= \langle \delta b, [\Sigma]\rangle=\langle b, \partial [\Sigma]\rangle=0 \, .
\end{equation}
However, on manifolds with boundary $\langle b, \partial [\Sigma]\rangle\in \mathbb{Z}_N$, and the weighting factors are not necessarily gauge-invariant or independent of the connection on the bundle. In general, the non-invariance of the ``topological" term is completely controlled by the connection on the boundary. Honest gauge transformations that vanish at the boundary will not change the path integral. However, if we consider a gauge transformation that does not vanish on some $S^1$ boundary, and instead describes a nontrivial element of $\pi_1(G)$ as we wind around the boundary circle, then $\langle b, \partial[\Sigma]\rangle$ will be nonzero and the path integral will transform by a phase.

The boundary values of the gauge fields represent background gauge fields for the global symmetry $G$ in the random matrix ensemble. A gauge transformation that does not vanish on the boundary represents a gauge transformation of this background gauge field. The non-invariance of the partition function under such a transformation is the definition of an 't Hooft anomaly. 

With this understanding, let's now try to define the partition function with boundaries in the theory with the anomaly. The naive generalization of \eqref{eq:aksigma2} to a manifold with a single boundary and prescribed holonomy $U\in G$ would be
\begin{equation}\label{eq:anomPart}
    Z_\Sigma^{(k)}(\beta_1,U_1) \stackrel{?}{=}\sum_{\omega=0}^{N-1}e^{2\pi \i k\omega/N}\; ``Z_{\Sigma,\,\omega}(\beta_1,U_1)"\; .
\end{equation}
In this formula, one would like $``Z_{\Sigma,\,\omega}(\beta_1,U_1)"$ to denote the partition function of the $G$ gauge theory restricted ``to the class of bundles with topological number  $\omega$ and prescribed holonomy $U$". As we have noted, this formal object is ill-defined, since the value of $\omega$ on a manifold with boundary is no longer an invariant. The left hand side of the equation is well-defined but not invariant under large gauge transformations that wind non-trivially around the boundary circle. These gauge transformations correspond to non-trivial elements of $\pi_1(G)$ and change the phase of the path integral in a controlled way. This phase ambiguity can be repackaged by lifting the holonomy $U_1$ to an element $\tilde{U}_1$ of $\pi^{-1}(U_1)\in \widetilde{G}$. Since there are no topologically non-trivial gauge transformations in $\widetilde{G}$, once the lift of $U_1$ is chosen the path integral is defined unambiguously. However, different choices of the lift result in different overall phases, matching the ambiguity in the $G$-theory's partition function. Having chosen a definite phase for the path integral by lifting to $\widetilde{G}$, the partition functions in the fixed topological sectors can be computed by excising a disk from $\Sigma$ and gluing in one with holonomy $u\in \mathbb{Z}_N$, as in Section \ref{secbundleintro}. A well-defined version of \eqref{eq:anomPart} therefore reads
\begin{equation}
      Z^{(k)}_\Sigma(\beta_1,\widetilde{U}_1 ) = N^{-2g}\sum_{u \in \mathbb{Z}_N}
  e^{2\pi\i k \omega(u)/N} \, \widetilde{Z}_{\Sigma}(\beta_1,\widetilde{U}_1; u)\, .
\end{equation}
Again, the $N^{-2g}$ factor can be interpreted as the inverse of a $\mathbb{Z}_N$ partition function on a genus $g$ surface with one boundary.

The discussion is identical on a surface with $n$ boundaries. One would like to compute the partition function
\begin{equation}
    Z^{(k)}_g(\beta_1,U_1; \ldots;\beta_n, U_n)
\end{equation}
with specified honolomy $U_i \in G$ on the $i^\text{th}$ boundary component. This object has phase ambiguities associated to elements of $\pi_1(G)$ acting at each boundary. These ambiguities can be repackaged by working with a $\widetilde{G}$-theory and lifting the $U_i$ to elements of the covering group $\widetilde{U}_i\in\pi^{-1}(U_i)\in \widetilde{G}$. There is an $N$-fold choice for each $\widetilde{U}_i$, and there is no canonical way to make this choice.\footnote{Note that if $U_i=1_G$, then there is a natural choice $\widetilde{U}_i=1_{\widetilde{G}}$, since the identity element of every group is unique.} In this formulation, the dependence of the partition function on the choice of an element of $\pi^{-1}(U_i)$ is the statement of the anomaly. The phase ambiguities match those of the $G$-theory: the partition function transforms the same way (by a phase related to an element of the center of $\widetilde{G}$) when one performs a large gauge transformation in the $G$ theory or when one changes the choice of lift in $\pi^{-1}(U_i)$ in the $\widetilde{G}$-theory.
Following the prescription of excising a small disk in \cite{Witten:1991we}, we get the analog of (\ref{eq:zsigmausun}) and (\ref{eq:aksigma2}) for general manifolds with boundary:
\begin{align}
  Z^{(k)}_g(\beta_1,\widetilde{U}_1; \ldots; \beta_n,\widetilde{U}_n) &= N^{1-2g-n}\sum_{u \in \mathbb{Z}_N}
  e^{2\pi\i k \omega(u)/N} \, \widetilde{Z}_{g}(\beta_1,\widetilde{U}_1; \ldots; \beta_n,\widetilde{U}_n; u)\\
  &=\sum_{\widetilde{r}}\delta_{k,k'(\widetilde{r})}\dim(\widetilde{r})^n\left(\frac{\vol(G)}{\dim(\widetilde{r})}\right)^{2g+n-2}\prod_{j = 1}^n e^{-\beta_jc_2(\widetilde{r})/2}\frac{\chi_{\widetilde{r}}(\widetilde{U}_j)}{\dim(\widetilde{r})}.
\label{nboundariesanomaly}
\end{align}
The Kronecker delta ensures that we sum only over representations of $N$-ality $k$. In the case of a finite group, we would have the same formula, with $\vol(G)$ replaced by $|G|$.

Let's see how this formula reproduces the 't Hooft anomaly. Suppose that we transform one of the holonomies $U_i$, by a background gauge transformation that corresponds to the elementary generator of $\pi_1(G)$. $U_i$ will just transform by conjugation, but the lift $\widetilde{U}_i$ will be multiplied by $z$, the generator of $\ker(\pi)$. Using (\ref{nality}), we find that (\ref{nboundariesanomaly}) will transform by a definite phase $e^{2\pi \mathrm{i}k/N}$ which characterizes the 't Hooft anomaly. In particular, in the case $k = 0$, we sum over honest linear representations, and the result is just the partition function in the theory with no anomaly.

\subsection{Connection to random matrix theory}
\label{sec:impanom}
The formula (\ref{nboundariesanomaly}) (or its analog for finite groups) can be converted to the representation basis by exploiting the orthogonality of characters of $\widetilde{G}$, and the projection formula (\ref{projectiontor}) for the group $\widetilde{G}$.\footnote{After projecting (\ref{nboundariesanomaly}) to a given $\widetilde{r}$ using the orthogonality of the characters of $\widetilde{G}$, we are only left with a dependence on $\dim(\widetilde{r})$. Thus the answer is independent of the choice of $\widetilde{G}$, since the various possible $\widetilde{G}$ are isoclinic.} The resulting expression for $Z_g(\beta_1, r_1; \ldots; \beta_n, r_n)$ is identical to (\ref{BFans}), except for the restriction to projective representations of fixed $N$-ality. The discussion of section \ref{sec:comparison} goes through as before, and we conclude that the bulk theory is dual to a random matrix ensemble with $G$ symmetry but only projective representations of fixed $N$-ality.

\subsection{Examples}
In the rest of this section, we consider specific examples to illustrate the above general theory.
In section \ref{secztwotwo}, we consider the finite group $G=\ztwotwo$, for which there are two Schur covering groups: the dihedral group with eight elements and the quaternion group.
In section \ref{secotwo}, we consider the case with $G = O(2)$ and $\widetilde{G} = \pin(2)$.
Here $G$ is continuous, but disconnected.
In section \ref{secpsun}, we consider the case $G=PSU(N)$ and $\widetilde{G} = SU(N)$.
Here $G$ is connected, but not simply connected, and $\widetilde{G}$ is its universal cover; this is the original setting in \cite{Witten:1978ka}.
\subsubsection{$\mathbb{Z}_2\times \mathbb{Z}_2$}
\label{secztwotwo}

The simplest internal symmetry group capable of exhibiting an anomaly in quantum mechanics is $\mathbb{Z}_2 \times \mathbb{Z}_2$, see the second column of Table 1 of \cite{Chen:2011pg}. Since $H^2(\mathbb{Z}_2 \times \mathbb{Z}_2,U(1))=\mathbb{Z}_2$, there are two choices for the gauge theory action and a single non-trivial class of projective representation. However, there are multiple inequivalent central extensions of $\mathbb{Z}_2\times \mathbb{Z}_2$ by $\mathbb{Z}_2$ classified by the group $H^2(\mathbb{Z}_2\times \mathbb{Z}_2,\mathbb{Z}_2)=\mathbb{Z}_2\oplus\mathbb{Z}_2\oplus\mathbb{Z}_2$.
There are four possible $\widetilde{G}$'s, and thus multiple elements of $H^2(\mathbb{Z}_2\times \mathbb{Z}_2,\mathbb{Z}_2)$ can correspond to the same $\widetilde{G}$. 
Explicitly, there is $\mathbb{Z}_2 \times \mathbb{Z}_2 \times \mathbb{Z}_2$ (appearing once), $\mathbb{Z}_2 \times \mathbb{Z}_4$ (appearing thrice), the dihedral group $D_8$ (appearing thrice), and the quaternion group $Q$ (appearing once).
Of these, $D_8$ and $Q$ are isoclinic Schur covering groups, while $\mathbb{Z}_2\times \mathbb{Z}_2 \times \mathbb{Z}_2$ and $\mathbb{Z}_2\times \mathbb{Z}_4$ are not stem extensions, and therefore are not Schur covering groups.\footnote{A discussion of these facts can be found \href{https://groupprops.subwiki.org/wiki/Second_cohomology_group_for_trivial_group_action_of_V4_on_Z2\#Elements}{here} or in the lecture notes \cite{Greggroup} by Greg Moore.}

Since $\mathbb{Z}_2\times \mathbb{Z}_2$ is abelian, it suffices to work with the abelianization of $\pi_1(\Sigma)$ when enumerating bundles. 
Each bundle is characterized by a homomorphism $H_1(\Sigma,\mathbb{Z}) \to \mathbb{Z}_2\times \mathbb{Z}_2$, of which there are $4^{2g}$. The simplest theory with this gauge symmetry counts the number of such bundles modulo the conjugation action, and has the partition function
\begin{equation}
    Z_g^{(k=0)} =\frac{1}{|\mathbb{Z}_2\times \mathbb{Z}_2|}\sum_{k=1}^{4^{2g}} \;1= 4^{2g-1}\, , \qquad 
    \text{($\ztwotwo$ theory without anomaly)}\,  .
\label{eq:zztwotwo}
\end{equation}
This expression in the ``bundle basis" is of course also equal to the expression in the representation basis \eqref{Zclosed}. The group $\mathbb{Z}_2\times \mathbb{Z}_2$ has 4 conjugacy classes and 4 one-dimensional representations, so the partition function can also be expressed as
\begin{equation}
    Z_g^{(k=0)} =\sum_{r}\left(\frac{|\mathbb{Z}_2\times \mathbb{Z}_2|}{\dim (r)}\right)^{2g-2}=4\cdot 4^{2g-2} \; .
\end{equation}

 It is possible to construct a variant of this theory that distinguishes the $\mathbb{Z}_2\times\mathbb{Z}_2$ bundles that lift to bundles of the covering group $D_8$ from those that do not.\footnote{ 
 See appendix D of \cite{Gaiotto:2017yup} for a recent discussion.
 As mentioned previously, the $\mathbb{Z}_2\times \mathbb{Z}_2$ bundles can be distinguished using either Schur covering group $D_8$ or $Q$. Both groups are isoclinic and therefore have identical partition functions on closed orientable surfaces. } The corresponding action is related to the non-trivial element of $H^2(\mathbb{Z}_2\times \mathbb{Z}_2,U(1))$.  The dihedral group has the presentation 
 \begin{equation}
     D_8=\langle \;a,b \;|\; a^4=1,\; b^2=1,\; bab^{-1}=a^{-1} \;\rangle.
 \end{equation}
 The elements $\{1,a^2 \}$ are central, and $D_8/\{1,a^2\} = \ztwotwo$.
The group $D_8$ has five conjugacy classes, and therefore five irreducible representations.
Four of these representations are one-dimensional and restrict to representations of $\ztwotwo$ (since $a^2$ is represented trivially).
The fifth irreducible representation of $D_8$ has dimension two, and the central element $a^2$ is represented by $-I_{2\times 2}$. 
This corresponds to the unique class of non-trivial projective representation of $\ztwotwo$.
The path integral of the $D_8$ gauge theory on a closed surface is given by the sum over representations \eqref{Zclosed}
\begin{equation}\label{ZD8}
\widetilde{Z}_g = |D_8|^{2g-2} \sum_{\tilde{r}}\text{dim}(\tilde{r})^{2-2g}=8^{2g-2}(4 +2^{2-2g})\, .
\end{equation}

We will now explicitly evaluate the partition functions of the $\mathbb{Z}_2\times \mathbb{Z}_2$ theory with the anomaly. 
The topological action \eqref{topAct} corresponding to the non-trivial element of $H^2(\mathbb{Z}_2\times \mathbb{Z}_2,U(1))=\mathbb{Z}_2$ evaluates to $0$ if the $\mathbb{Z}_2\times \mathbb{Z}_2$ bundle lifts to a $D_8$ bundle, and returns $\frac12$ otherwise.
The partition function with the anomaly therefore takes the form
\begin{equation}\label{Zdiscrete}
    Z_g^{(k=1)} =  \frac{1}{|\mathbb{Z}_2\times \mathbb{Z}_2|}\sum_{\gamma} e^{\i \pi \omega(\gamma)}= \sum_{\omega=0,1} (-1)^\omega Z_{g, \, \omega} \, , \qquad 
    \gamma \in \text{Hom}(\pi_1(\Sigma),\mathbb{Z}_2\times \mathbb{Z}_2)\, ,
\end{equation}
where $\omega(\gamma)=0$ if $\gamma$ lifts to a $D_8$ bundle and $\omega(\gamma)=1$ otherwise.
The calculation therefore reduces to counting $\mathbb{Z}_2\times \mathbb{Z}_2$ bundles with signs. The sum over $\mathbb{Z}_2\times \mathbb{Z}_2$ bundles that lift to $D_8$ bundles is simply related to the $D_8$ partition function \eqref{ZD8} in the representation basis through the formula \eqref{eq:zomegasigma} and gives
\begin{align}
    Z_{g, \, \omega=0}= \frac{4^{2g-2}}{2}  \left(1 + 1 + 1 + 1 + 2^{2-2g} \right)\, .
\label{eq:zomegazero}
\end{align}
Because the kernel of the covering map $D_8\to \mathbb{Z}_2\times \mathbb{Z}_2$ contains a single non-identity element $a^2$, the non-lifting bundles can all be constructed from the liftable bundles by excising a small disk and imposing the holonomy $a^2$ around it. According to the general formula \eqref{eq:zomegasigma},
each $D_8$ representation occurring in this second sum is weighted by plus or minus one depending on whether $a^2$ is represented nontrivially in that representation. One easily obtains
\begin{align}
    Z_{g, \, \omega=1} = \frac{4^{2g-2}}{2} \left(1 + 1 + 1 + 1 - 2^{2-2g} \right)\, .
\label{eq:zomegaone}
\end{align}
The partition function of the theory without the anomaly is the sum of (\ref{eq:zomegazero}) and (\ref{eq:zomegaone}), also known as $4^{2g-1}$. This is in complete agreement with (\ref{eq:zztwotwo}). Note that the sum projects out the two-dimensional representation of $D_8$ which defines a genuinely projective representation of $\mathbb{Z}_2\times \mathbb{Z}_2$.
The partition function of the theory with the anomaly, Eq. (\ref{Zdiscrete}), is the \emph{difference} of (\ref{eq:zomegazero}) and (\ref{eq:zomegaone}), and projects onto the single genuinely projective representation:
\begin{align}
    Z^{(k=1)}_g &= Z_{g,\omega=0} - 
    Z_{g,\omega=1} \\
    &= 4^{g-1}\, .
\end{align}
We emphasize that if one is simply counting the number of bundles, and not quotienting by conjugations, one is computing the volume of moduli space rather than the partition function.
The moduli space volumes for a finite group are simply equal to $|G|$ times the relevant partition function.

It is instructive to check these results on the torus.
There are $8^2 = 64$ possible assignments of $D_8$ holonomies $(u,v)$ to the two homology cycles of the torus.
Since $D_8$ is nonabelian, the necessary condition $uvu^{-1}v^{-1}=1$ for a $D_8$ bundle is not automatically satisfied, and only 40 choices of $(u,v)$ satisfy the relation and define $D_8$ bundles.
These 40 bundles, after accounting for conjugation, yield $\widetilde{Z}_{g=1} = 5$ in agreement with (\ref{ZD8}) for $g=1$.
The remaining $24$ pairs $(u,v)$ satisfy $uvu^{-1}v^{-1}=a^2$ and are accounted for in (\ref{eq:zomegaone}).
In the $\ztwotwo$ language, out of the sixteen possible assignments  of $\mathbb{Z}_2\times \mathbb{Z}_2$ holonomy $(U,V)$  to the two cycles of the torus, ten pairs lift to admissible $D_8$ holonomies, while six do not.

We now turn to the evaluation of the $\mathbb{Z}_2\times \mathbb{Z}_2$ gauge theory partition function on surfaces with boundary. In the theory without the anomaly, we need to specify a holonomy $U_i \in \ztwotwo$ on each boundary $i \in \{1, \ldots, n\}$. Since the discrete gauge theories are truly topological and independent of the length of the boundary curves, we can simply use the answers from \cite{Witten:1991we}:
\begin{equation}
    Z_{g}^{(k=0)}(\beta_1,r_1;\dots;\beta_n,r_n) = 4^{2g+n-2}\, \delta_{r_1, \dots, r_n} .
\end{equation}

Evaluation of this partition function in the theory with the anomaly is more subtle. 
As discussed in section \ref{sec:anombound}, we need to specify a holonomy $U_i \in \ztwotwo$ for each boundary, and also a choice of lift $\widetilde{U}_i \in \pi^{-1}(U_i) \subset D_8$. Because the lifts of the non-identity elements of $\mathbb{Z}_2\times \mathbb{Z}_2$ are traceless in the two-dimensional representation of $D_8$, the answer for the path integral in the theory with the anomaly should vanish unless all of the $U_i$'s are equal to the identity. 

Let's check this mechanically. Suppose that we glue a trumpet with holonomy $U$ around the $S^1$ into a larger manifold. There is a choice of Wilson line $V$ across the $S^1$ gluing locus, and we need to sum over this. The fact that this sum gives zero can be established using the locality argument in \cite{Stanford:2019vob}: if we zoom in on a neighborhood of the gluing, then we will find a geometry with the topology of a cylinder $S^1\times I$, with holonomy $U$ on the $S^1$ and Wilson line $V$ on the interval. This same sum arises in the calculation of the torus partition function with fixed holonomy $U$ around one of the cycles, and arbitrary holonomy $v$ around the other.

We therefore consider the restricted holonomy sum on the torus, where $U = a$, and we sum over $V = 1,a,b,ab$. Because $\mathbb{Z}_2\times\mathbb{Z}_2$ is abelian, we will have $UVU^{-1}V^{-1} = 1$ for any of these four choices. To check whether the bundle lifts to $D_8$, we need to check if $UVU^{-1}V^{-1}=1$ using only the relations of $D_8$. One can check
\begin{equation}
    a 1 a^{-1} 1^{-1} = 1, \hspace{20pt} a a a^{-1} a^{-1} = 1, \hspace{20pt} a b a^{-1} b^{-1} = a^2, \hspace{20pt} a(ab)a^{-1}(ab)^{-1} = a^2.
\end{equation}
In $\mathbb{Z}_2\times \mathbb{Z}_2$, we have $a^2 = 1$, but in $D_8$, it is not one, so one finds that half of the bundles lift and half do not. So the conclusion is that on the torus with holonomy on one cycle restricted to be $U = a$, the sum over gauge bundles weighted by the anomaly term vanishes. One can repeat this calculation for the case $U = b$ or $U = ab$ with the same conclusion.

\subsubsection{$O(2)$}
\label{secotwo}
We now consider the simplest example of a continuous, but disconnected symmetry group, $O(2)=U(1) \rtimes \mathbb{Z}_2$.
We denote the connected component of the identity in $O(2)$ by $O(2)^+$, and the other connected component by $O(2)^-$.
An $O(2)$ symmetry naturally arises in systems with a $U(1)$ symmetry that are invariant under charge conjugation, and the possible mixed anomaly between $U(1)$ and $\mathbb{Z}_2$ has interesting consequences both in quantum mechanics and in gauge theory \cite{Gaiotto:2017yup, Cordova:2019jnf}.\footnote{Certain SYK models with $N$ Dirac fermions exhibit $O(2)$ symmetry.
In the SYK context, the anomaly is present when $N$ is odd, and it is absent if $N$ is even \cite{Stanford:2017thb, Fu:2016vas}.}

By an $O(2)$ BF gauge theory, we mean a sum over all flat $O(2)$ bundles. 
There are several classes of $O(2)$ bundle, each distinguished by a characteristic class. 
The first Stiefel-Whitney class $\omega_1\in H^1(\Sigma,\mathbb{Z}_2)$ is an obstruction to orientability of the bundle. 
If it vanishes, the $O(2)$ bundle defines an $SO(2)$ bundle, or a complex line bundle. 
Such bundles are classified by their first Chern class $c_1\in H^2(\Sigma,\mathbb{R})$ which vanishes for a flat bundle. 
If $\omega_1\neq 0$, then the Chern class cannot be defined for the bundle, but a mod $2$ analog $\omega_2\in H^2(\Sigma,\mathbb{Z}_2)$, the second Stiefel-Whitney class, still makes sense. 
Bundles with $\omega_1=\omega_2=0$ can be lifted to $\text{Spin}(2)$ bundles, and 
bundles with $\omega_2=0$ can be lifted to $\text{Pin}^+(2)$ bundles (whether or not $\omega_1$ is trivial).
Since we are not considering matrix ensembles with time reversal symmetry, all Riemann surfaces in this section are oriented.
For rank-2 vector bundles over oriented Riemann surfaces $\omega_1\cup \omega_1=0$,
and so the only topological weighting factor for $O(2)$ bundles that is consistent with factorization is $\omega_2$. 
There is thus a unique theory with anomalous $O(2)$ symmetry \cite{Chen:2011pg}, and it corresponds to weighting the $O(2)$ bundles according to whether or not they admit a $\pin(2)$ lift.\footnote{
Since $\omega_1\cup \omega_1=0$ on oriented surfaces, $\omega_2$ is the obstruction for both Pin$^+(2)$ and Pin$^-(2)$. 
We could therefore also use the group $\text{Pin}^-(2)$ in the following discussion. 
Since Pin$^+(2)$ and Pin$^{-}(2)$ are isoclinic, their partition functions on closed surfaces agree and the choice does not effect the calculation of the $O(2)$ partition functions. 
This is completely analogous to the choice of $D_8$ or $Q$ in section \ref{secztwotwo}.}

The group $\pin(2)$ is a double cover of $O(2)$, with the kernel of the covering map given by the center of $\pin(2)$:
\begin{align}
    1 \to \mathbb{Z}_2 \to \pin(2) \to O(2) \to 1\, .
\end{align}
Not all $O(2)$ bundles lift to $\pin(2)$ bundles, and not all representations of $\pin(2)$ descend to $O(2)$ representations. 
The group $\pin(2)$ has two one-dimensional irreducible representations: the trivial representation and the determinant representation, both of which restrict to representations of $O(2)$. 
In addition, there are an infinite number of two-dimensional representations labelled by elements of $m\in \mathbb{N}/2$.
The representations with integer $m$ descend to representations of $O(2)$.
When $m$ is not an integer, the nontrivial central element of $\pin(2)$ is represented by $-I_{2\times 2}$, and the $\pin(2)$ representation defines a genuinely projective representation of $O(2)$.
Following the general discussion in Section \ref{secbundleintro}, we expect the $O(2)$ symmetric quantum mechanical system with the anomaly to contain precisely these projective representations.

We first consider the theory without the anomaly, where the general formulas of section \ref{sec:BF} are valid.
The partition function on a closed genus-$g$ surface is given by the sum over all irreducible representations, as in  (\ref{Zclosed}): 
\begin{equation}
Z_{g}^{(k=0)} = \left(4\pi\right)^{2g-2}
\left(1+1+\sum_{m\in \mathbb{N}} 2^{2-2g}\right) \, ,
\label{o2closed}
\end{equation}
where the factor of $4\pi$ in this expression arises because $\vol(O(2)) = 4\pi$, in a standard convention.
Note that this partition function (\ref{o2closed}) is formally divergent, because we have infinitely many 2d irreps and no Boltzmann suppression on closed surfaces. 
For Riemann surfaces with boundary, the partition functions in the representation basis are given by the general expression (\ref{BFans}).
In the holonomy basis, we will have
\begin{align}
    Z_g^{(k=0)}(\beta_1,u_1; \ldots, \beta_n,u_n) = 
    (4\pi)^{2g+n-2} 
    \left(
    1 + \prod_{i=1}^n\text{det}(u_i) + 
    2^{2-2g-n}
    \sum_{m \in  \mathbb{N}}
    \prod_{i=1}^n (2\cos m \theta_i) 
    e^{-\beta_i m^2/2}
    \right)\, .
\label{eqo2bdyznoanom}
\end{align}

Now, let us turn to the theory with the anomaly.
As discussed above, the anomaly term is the second Stiefel-Whitney class of the $O(2)$ bundle \cite{Komargodski:2017dmc}.\footnote{In more geometric terms, for a rank-2 vector bundle $E$ over a Riemann surface $\Sigma$, the quantity $\int_\Sigma \omega_2$ is the obstruction to having a section of unit vectors.}
We can compute the partition function on closed surfaces using the general method described in section \ref{secbundleintro}. The results are
\begin{align}
    Z_{g, \,\omega=0} &= \frac{1}{2} (4\pi)^{2g-2} \left( 1 + 1 + \sum_{m \in \{\frac{1}{2}, 1, \frac{3}{2} \ldots \}}
    2^{2-2g}  \right) \, , \\
    Z_{g,\, \omega=1} &= \frac{1}{2} (4\pi)^{2g-2} \left( 
    1 + 1 + \sum_{m \in \{\frac{1}{2}, 1, \frac{3}{2} \ldots \}}
    2^{2-2g} (-1)^{2m} 
    \right) \, .
\end{align}
Note that these expressions are also divergent.
Now we combine the above two equations using (\ref{eq:aksigma2}) to get the formal partition function
\begin{align}
    Z_{g}^{(k=1)} &= Z_{g,\omega=0} - Z_{g, \omega=1}
    = (4\pi)^{2g-2}  \sum_{m \in \{\frac{1}{2}, \frac{3}{2} \ldots \}}
    2^{2-2g} \, .
\end{align}

Next we consider manifolds with boundary in the theory with the anomaly.
The partition function vanishes if any of the boundary holonomies belong to $O(2)^-$,  since these elements are represented by traceless elements in the projective representations.
If a boundary holonomy $U$ belongs to $O(2)^+$, then the phase of the partition function is ambiguous and we have to specify a lift of this holonomy to $\pin(2)$, as discussed in section \ref{sec:anombound}.
This lift requires a choice since there are two elements $\widetilde{U}, \widetilde{U}' \in \pin(2)$ that map to $U\in O(2)$ under the covering map $\pi: \pin(2) \to O(2)$.
There is no canonical choice of $\widetilde{U}$ or $\widetilde{U}'$, and 
the partition function flips sign under the interchange. To see this explicitly, parametrize the connected component of the identity in $\pin(2)$ by an angle $\theta$ which has periodicity $4\pi$.
For fixed $\widetilde{U}_1, \ldots \widetilde{U}_n \in \pin(2)$, the $n$-boundary partition function takes the form (\ref{nboundariesanomaly}), which in this case is explicitly given as
\begin{align}
    Z_g^{(k=1)}(\beta_1,\widetilde{U}_1; \ldots, \beta_n,\widetilde{U}_n) = 
    (2\pi)^{2g+n-2} \sum_{m \in \{\frac{1}{2}, \frac{3}{2} \ldots \}}
    \prod_{i=1}^n (2\cos m \theta_i)
    e^{-\beta_i m^2/2} \, .
\label{eqo2bdyz}
\end{align}
Changing the lift of the holonomy from $\widetilde{U}_i$ to $\widetilde{U}_i'$ amounts to sending $\theta_i\to \theta_i+2\pi$.
Since all of the $m$'s appearing in the sum are of the form $(\text{integer}+1/2)$, the partition function flips sign under the shift  of $\theta_i$ by $2\pi$.
This is one of the defining features of the theory with the anomaly, and we see it explitly in (\ref{eqo2bdyz}). 
We can also rewrite this in the representation basis, as we explained in section \ref{sec:impanom}.
Note also the the partition functions on manifolds with a boundary are convergent, as opposed to closed surfaces, because of the presence of Boltzmann suppression factors.

\subsubsection{$PSU(N)$}
\label{secpsun}
The last example that we will consider is $PSU(N)$ gauge theory. 
$PSU(N)$ is connected but not simply connected, and  can be viewed either as the quotient of $SU(N)$ by its center $\mathbb{Z}_N$, or as the quotient of $U(N)$ by its center.

On a closed, oriented surface there is a single topological class of $SU(N)$ principal bundle. In order to construct a non-trivial $PSU(N)$ bundle from this $SU(N)$ bundle, one excises a disk and glues in one with holonomy ${u\in \mathbb{Z}_N}$. Said differently, one relaxes the $SU(N)$ cocycle condition on the transition functions $g_{ij}$ to require
$g_{ij}g_{jk}g_{kl}\in \mathbb{Z}_N$,
 since these elements are represented trivially under the projection ${\pi:SU(N)\to PSU(N)}$. The assignment of elements of $\mathbb{Z}_N$ to each triple overlap on the surface defines a class  $[\omega_N]\in H^2(\Sigma,\mathbb{Z}_N)$ through the \v{C}ech cohomology, and the pairing of this class with the fundamental homology class defines a topological invariant ${\omega=\langle [\omega_N], [\Sigma]\rangle\in \mathbb{Z}_N}$ of the bundle. This number can be interpreted as the mod $N$ first Chern class of a corresponding $U(N)$ bundle.
The possible theories are labeled by an integer ${k\in \{0,\dots N-1\} }$ and the partition functions take the form
  \begin{equation}
      Z^{(k)}_\Sigma =\sum_{\omega=0}^{N-1}e^{2\pi \i k\omega/N} Z_{\Sigma,\,\omega}\, .
  \end{equation}
For example, in the case $N=2$ for which ${PSU(2)=SO(3)=SU(2)/ \{\pm 1 \} }$, there are exactly two classes of $SO(3)$ bundle.  The relevant $SO(3)$ partition functions on a genus $g$ surface are 
 \begin{equation}
Z_{g,\omega=0} = \frac{1}{2}\text{vol}(SO(3))^{2g-2}\sum_{n\in \mathbb{N}}\frac{1}{n^{2g-2}} \, , \hspace{.75 in}   Z_{g, \,\omega=1}=\frac{1}{2}\text{vol}(SO(3))^{2g-2}\sum_{n\in \mathbb{N}}\frac{(-1)^{n+1}}{n^{2g-2}} \, .
\end{equation}
The partition function of the $SO(3)$ theory without the anomaly is
  \begin{equation}
      Z^{(k=0)}_g = Z_{g,\omega=0}+Z_{g,\omega=1}=\text{vol}(SO(3))^{2g-2}\sum_{n\in 2\mathbb{N}+1}\frac{1}{n^{2g-2}} \, ,
  \end{equation}
  and contains only the genuine linear representations of $SO(3)$.
The $SO(3)$ theory with the anomaly has
  \begin{equation}
      Z^{(k=1)}_g = Z_{g,\omega=0}-Z_{g,\omega=1}=\text{vol}(SO(3))^{2g-2}\sum_{n\in 2\mathbb{N}}\frac{1}{n^{2g-2}}\, ,
  \end{equation}
and contains only the genuinely projective representations of $SO(3)$, as described in Section \ref{secbundleintro}. On a surface  with boundary but trivial holonomy, the partition functions take a similar form with Boltzmann factors and the appropriate classes of representations.

The topological term requires more care on manifolds with boundary and non-trivial holonomy (corresponding to non-trivial background fields in the quantum mechanics). 
The class $[\omega_N]$ is a topological invariant of the bundle because under a change of connection, the representative changes by a coboundary:
\begin{equation}
      \omega_N[A]-\omega_N[A']=\delta b, \;\;\;\; b \in C^1(\Sigma,\mathbb{Z}_N)\, .
\label{changeofgauge}
\end{equation}
On a manifold without boundary ${\langle \delta b ,[\Sigma]\rangle=0}$ so that the invariant is independent of any choices that went into its definition. 
On a manifold with boundary $\partial \Sigma$, one needs to specify both boundary conditions on $\partial \Sigma$, and in order to evaluate the topological term one also has to specify a representative of the cohomology class $\omega_N$.
If one simply specifies a holonomy $U\in PSU(N)$ on $\partial \Sigma$, the phase of the partition function transforms under large gauge transformations that do not vanish at the boundary circle (non-trivial elements of $\pi_1(PSU(N))=\mathbb{Z}_N$)
\begin{equation}
      Z^{(k)}\to 
      \exp \left( 2\pi \i \, \frac{k}{N} 
      \langle b , \partial \Sigma \rangle \right)\, Z^{(k)}(\omega_N[A])\, ,
\end{equation}
where $b$ is the 1-cocycle defined in (\ref{changeofgauge}).
The explicit formulas for partition functions can be written using the methods of section \ref{sec:anombound}.

\section{Incorporating time-reversal symmetry and unorientable surfaces}
\label{sec:timereversal}

So far, we have only discussed the gauge theory on orientable surfaces. 
However, one can also consider a bulk theory in which orientation-reversal is gauged. 
Concretely, this means that surfaces can be glued together with a reversal of orientation, and so in particular, unorientable surfaces should be allowed. 
Bulk theories in which orientation is gauged are expected to be related to boundary theories with time-reversal symmetry. 
In the context of random matrix theory, this would mean GOE-like or GSE-like matrix ensembles, see  \cite{Harris:1990kc,Neuberger:1980qh} in the ``minimal string'' and \cite{Stanford:2019vob} in JT gravity.

Following the strategy of the first parts of this paper, we will imagine starting with a ``seed'' unoriented 2d gravity theory such that $Z^{\text{grav}}(\beta_1\dots\beta_n)$ agrees with the genus expansion of a GOE-like or GSE-like random matrix integral. 
Then we will ask what happens when we include a gauge theory for the group $G$.

As a first step, let's discuss the gluing rule in the gauge theory when we glue two manifolds together across a $S^1$ boundary with an orientation reversing twist. 
It is convenient to start with the ordinary gluing rule (\ref{gluingrule}) in the holonomy basis,
\begin{equation}
    \int\frac{\mathrm{d}U}{\vol(G)}\psi(U^{-1})\phi(U).
\end{equation}
Reversing the orientation of one of the boundaries prior to gluing  replaces $U\leftrightarrow U^{-1}$ in one factor, leading to \cite{Witten:1991we}
\begin{equation}
    \int \frac{\mathrm{d}U}{\vol(G)}\psi(U)\phi(U) = \sum_r \psi(\overline{r})\phi(r).\label{gluingOr}
\end{equation}
In the final expression we wrote the answer in the representation basis, using the property of the characters $\chi_r(U^{-1}) = \chi_{\overline{r}}(U)$. 
The new gluing rule is simple in the representation basis, and glues representation $r$ to its conjugate representation $\overline{r}$.

The other new ingredient we need to consider is the possibility of including unorientable surfaces in the path integral. 
These surfaces can be constructed by cutting $n_c$ holes in an orientable surface of genus $g_0$, and attaching $n_c$ ``crosscaps.'' 
We use $g = g_0 + \frac{n_c}{2}$ to denote the genus of the surface after this operation. 
In order to perform this gluing in the gauge theory, one only needs to know that the wave function of the gauge theory on a crosscap in the representation basis is (see \cite{Witten:1991we}, section 2.4)
\begin{equation}
    \Psi_{\frac{1}{2}}(r) =s_r\delta_{r,\overline{r}}.\label{xcap}
\end{equation}
The $\delta_{r,\overline{r}}$ factor indicates that unorientable surfaces only contribute to path integrals with boundary conditions that include real or pseudo-real representations of the group $G$. The factor $s_r$ is one if the representation is real, and minus one if the representation is pseudo-real \cite{Witten:1991we}.

Note that in order to add a crosscap to a space with at least one boundary, we need to add a three-holed sphere, and then glue in a crosscap on one of the new boundaries. So in the representation basis, the total factor associated to adding a crosscap will be 
\begin{equation}
    \frac{\vol(G)}{\dim(r)}s_r\delta_{r,\overline{r}},
\end{equation}
where the first factor comes from the additional three-holed sphere.

Let's now try to interpret these changes to the bulk theory from the perspective of random matrix theory. 
Suppose first that $r \neq \overline{r}$. 
Then (\ref{xcap}) vanishes and crosscaps do not contribute, but the modified gluing rule (\ref{gluingOr}) allows us to glue $r$ to $\overline{r}$. 
This implies that the only effect of gauging orientation-reversal in the bulk theory is to replace the $\delta_{r_1,\dots,r_n}$ in eq.~(\ref{BFans}) with a $\delta$ symbol that is equal to one if the representations are equal or conjugate, and zero otherwise. 
So in any correlation function, an insertion of $Z(\beta,r)$ has the same effect as an insertion of $Z(\beta,\overline{r})$. 
Thus $Z(\beta,r) = Z(\beta,\overline{r})$ identically. 
In random matrix theory, the interpretation of this statement is that the eigenvalues in sector $r$ remain GUE-like, but are precisely degenerate with corresponding eigenvalues in the sector $\overline{r}$.

Next, suppose that $r = \overline{r}$. Then crosscaps do contribute, and there are two different ways that any trumpet can be glued into a larger surface: with or without orientation reversal prior to gluing. 
The gauge theory does not notice the orientation reversal if $r = \overline{r}$. 
Also, the weighting of the crosscap $\vol(G)/\dim(r)$ is such that (\ref{BFans}) remains true on nonorientable surfaces (once we generalize $g\rightarrow g_0 + \frac{n_c}{2}$ where $g_0$ is the genus without crosscaps, and $n_c$ is the number of crosscaps). 
With this generalization of the genus, the random matrix prediction for GOE-like or GSE-like ensembles is also (\ref{matrixAgreement}), so if the bulk theory without the gauge theory was dual to a GOE-like or GSE-like ensemble, then after coupling to gauge theory, this remains true in sectors with $r = \overline{r}$. An interesting detail is that if $r$ is pseudo-real and not real, then the sign associated to a crosscap will be reversed from that of the parent theory, and the statistics will be changed from GOE-like to GSE-like and vice versa. (See \cite{MulaseWaldron} for a proof that this sign interchanges GSE and GOE.)

Both the $r = \overline{r}$ and $r \neq\overline{r}$ results  have a simple explanation in random matrix theory. We assume a time-reversal operator $\sf{T}$ that commmutes with the full Hamiltonian $H$. The operator $\sf{T}$ acts on the full Hilbert space in such a way that it maps $\mathcal{H}_r \leftrightarrow \mathcal{H}_{\overline{r}}$. If $r$ is real, then $\sf{T}$ preserves $\mathcal{H}_r$ and implies that the Hamiltonian restricted to this subspace commutes with $\sf{T}$. So the corresponding eigenvalues should be drawn from a time-reversal-invariant ensemble (GOE-like or GSE-like). On the other hand, if $r$ is complex, then $\sf{T}$ does not imply any time-reversal symmetry within $\mathcal{H}_r$. So the eigenvalues should remain GUE-like, but degenerate between $\mathcal{H}_r$ and $\mathcal{H}_{\overline{r}}$.

\subsection{An example of a mixed anomaly including time-reversal}\label{sec-time-rev-anom}
The inclusion of time-reversal symmetry makes new types of anomalies possible. 
We will illustrate this with the simplest example, where $G = \mathbb{Z}_2$. 
Without time reversal, there is no possible anomaly for this group. 
But with time-reversal, there is a mod 2 anomaly which we now discuss. 

The anomaly is described in the bulk theory by a new type of topological action for the sum over $\mathbb{Z}_2$ bundles, which would be trivial in a theory without orientation reversal. The term can be understood either as
\begin{equation}
\exp \left({\mathrm{i}\pi\int w_1^{\mathbb{Z}_2}\cup w_1^{\mathbb{Z}_2}} \right) 
\hspace{20pt} \text{or} \hspace{20pt} 
\exp \left({\mathrm{i}\pi \int w_1^{\mathbb{Z}_2}\cup w_1^{O(2)}} \right)\, .
\label{twofactors}
\end{equation}
As we will see, the two are equal.\footnote{A third option $e^{\mathrm{i}\pi\int w_1^{O(2)}\cup w_1^{O(2)}}$ is the same thing as $(-1)^{n_c}$ where $n_c$ is the number of crosscaps. This is related to an anomaly of a very simple kind,  turning a GOE-like matrix integral with $\sf{T}^2 = 1$ into a GSE-like one with $\sf{T}^2 = -1$.} 
Here, $w_1^{O(2)}$ is the first Stiefel-Whitney class of the tangent bundle. 
It assigns one to 1-cycles along which the orientation is reversed, and zero to 1-cycles along which the orientation is not reversed. 
Similarly, the class $w_1^{\mathbb{Z}_2}$ assigns one to 1-cycles with nontrivial $\mathbb{Z}_2$ holonomy, and zero to 1-cycles with trivial holonomy.

On an orientable 2-manifold, both expressions are equal to one and therefore trivial. 
For the second expression, this is obvious since orientability implies that $w_1^{O(2)} = 0$. 
To see that the first expression vanishes identically, recall that
the cup product algebra\footnote{For a discussion, see section 3.2 of \cite{Hatcher:478079}. } on an orientable Riemann surface is
\begin{equation}
    \alpha_i\cup \beta_j = \delta_{ij}\gamma \, ,
\end{equation}
where $\alpha_i$ and $\beta_j$ are the 1-cocycles dual to the standard $A_i$ and $B_j$ cycles of the Riemann surface, and $\gamma$ is the 2-cocycle. 
This algebra implies that  $w_1\cup w_1$ vanishes mod 2 for $w_1$ set equal to any integer combination of the $\alpha_i$ and $\beta_j$.

However, on an unorientable manifold, the expressions in (\ref{twofactors}) are in general nontrivial. 
An important special case is $\mathbb{RP}^2$, for which the cup product algebra with $\mathbb{Z}_2$ coefficients is
\begin{equation}
\alpha\cup\alpha = \gamma \, .
\end{equation}
Here $\alpha$ is dual to the 1-cycle $A$ along which the orientation is reversed,\footnote{If we view $\mathbb{RP}^2$ as a disk closed off by a crosscap, then $A$ is the cycle that goes ``through'' the crosscap.} and $\gamma$ is the 2-cycle. 
If we consider a $\mathbb{Z}_2$ bundle with nontrivial holonomy along $A$, then we will find
\begin{equation}
    w_1^{\mathbb{Z}_2}\cup w_1^{\mathbb{Z}_2} = w_1^{\mathbb{Z}_2}\cup w_1^{O(2)} = \alpha\cup\alpha = \gamma
\end{equation}
and therefore both expressions in (\ref{twofactors}) are equal to minus one. 
This has an important implication: it means that the sum over $\mathbb{Z}_2$ bundles on $\mathbb{RP}^2$ weighted by (\ref{twofactors}) vanishes. 
The answer for the trivial $\mathbb{Z}_2$ bundle is one, and the answer for the nontrivial bundle is minus one.

A general unorientable surface can be constructed by starting with an orientable surface, cutting holes, and gluing in crosscaps. 
Associated to each crosscap is a 1-cycle that passes through it, along which the orientation is reversed. 
The sum over $\mathbb{Z}_2$ bundles along this 1-cycle will vanish by the same argument as above. 
So we conclude that unorientable surfaces do not contribute to the path integral in the theory weighted by (\ref{twofactors}).

Although non-oriented surfaces do not contribute to the path integral, the theory with the weighting (\ref{twofactors}) does not simply reduce to the theory without orientation gauged. 
Gauging orientation allows the possibility to glue ``trumpets'' to the rest of the manifold with a reversal of orientation. 
In fact, locally at the point that a trumpet is glued to the rest of the manifold, there are two different $\mathbb{Z}_2$ options: one can reverse the orientation, and one can act with the nontrivial $\mathbb{Z}_2$ element. 
We need to understand the weighting of (\ref{twofactors}) for these four possibilities.

These four choices only affect the region near the gluing, and because the weighting factor (\ref{twofactors}) is local, we can study the sum over these four possibilities in a slightly simpler context, as explained in \cite{Stanford:2019vob}. 
Specifically, imagine that we are gluing two trumpets together to form the double-trumpet. 
Then we can glue the ``big ends'' of the double-trumpet together to form a closed manifold. 
In the case where the orientation is not reversed at the gluing, this manifold is topologically a torus. 
Since the torus is orientable, the weighting factor is trivial, and we learn that the weighting is one, regardless of whether or not we act with $\mathbb{Z}_2$ before gluing.

In the case where the orientation is reversed, gluing the ``big ends'' together gives a Klein bottle. The cup product algebra on the Klein bottle is as follows:
\begin{equation}
 \alpha\cup \alpha = \alpha\cup\beta = \gamma, \hspace{20pt}\beta\cup\beta = 0.
\end{equation}
Here $\alpha$ is dual to the 1-cycle that originally formed the boundary at the ``big end'' of the double trumpet, and $\beta$ is dual to the 1-cycle that goes through the double trumpet from one ``big end'' to the other. $\gamma$ is dual to the 2-cycle. 
For the tangent bundle, we have $w_1^{O(2)} = \beta$, since the orientation is reversed as we pass through the waist of the double trumpet. 
For the $\mathbb{Z}_2$ bundle, we have
\begin{equation}
w_1^{\mathbb{Z}_2} = \sigma \alpha + \tau \beta,
\end{equation}
where $\sigma,\tau$ are both either one or zero. 
We should sum over the two values of $\tau$, since this corresponds to gluing with or without an action of $\mathbb{Z}_2$. 
But we should hold $\sigma$ fixed, because this corresponds to the $\mathbb{Z}_2$ holonomy at the ``big end'' of the trumpet, which is fixed as part of the boundary conditions.
The result of the cup product is
\begin{equation}
     w_1^{\mathbb{Z}_2}\cup w_1^{\mathbb{Z}_2}  = w_1^{\mathbb{Z}_2}\cup w_1^{O(2)} = \sigma \gamma ,
\end{equation}
where we used $\sigma^2 = \sigma$. 
Because the answer does not depend on $\tau$, the weighting is the same for both choices of holonomy through the double trumpet, in other words for both choices of whether we act with $\mathbb{Z}_2$ before gluing together. 
However, we find that the weighting is one for the case where the holonomy at the ``big end'' is trivial, and the weighting is minus one in the other case. 

Adding this together with the contributions we computed using the torus, we find that the total answer vanishes in the case where the boundary holonomy is nontrivial, and the total answer is four in the case where it is trivial. 
The above discussion was for the double trumpet, but by locality a similar story will apply any time we glue a trumpet to any surface. 
In particular, the answer will vanish any time one of the trumpets has an asymptotic boundary with the nontrivial $\mathbb{Z}_2$ holonomy. 

Let's now give an interpretation of this in random matrix theory. 
We expect to find a correspondence to a matrix ensemble with a time-reversal symmetry $\sf{T}$ and a $\mathbb{Z}_2$ global symmetry, generated by an operator we will refer to as $\sf{Z}$. 
The Hilbert space breaks up into two sectors corresponding to the two representations of the $\mathbb{Z}_2$ symmetry, $\mathcal{H} = \mathcal{H}_+\oplus\mathcal{H}_-$, where $\sf{Z}$ acts as $\pm 1$ on the two subspaces. 
A trumpet boundary with trivial $\mathbb{Z}_2$ holonomy corresponds in the matrix integral to an insertion of
\begin{equation}
\tr_{\mathcal{H}}(e^{-\beta H}) =\tr_{\mathcal{H}_+}(e^{-\beta H}) + \tr_{\mathcal{H}_-}(e^{-\beta H}),
\end{equation}
while an insertion with nontrivial holonomy $\sf{Z}$ corresponds to an insertion of
\begin{equation}
    \tr( {\sf Z} e^{-\beta H}) = \tr_{\mathcal{H}_+}(e^{-\beta H}) - \tr_{\mathcal{H}_-}(e^{-\beta H}).
\end{equation}
The vanishing of all gravity path integrals with nontrivial holonomy means that $\tr( {\sf Z} e^{-\beta H})$ vanishes identically. 
This means that the two sectors $\mathcal{H}_-$ and $\mathcal{H}_+$ have precisely the same spectrum of energy eigenvalues. 
Note that because unorientable surfaces vanish, the statistics in either sector will be GUE-like.

This pattern can be explained by saying that the $\sf{Z}$ and $\sf{T}$ operators do not commute, but instead satisfy
\begin{equation}
    \sf{Z} \sf{T} = -\sf{T}\sf{Z}.
\end{equation}
The minus sign in this equation is the boundary expression of the mixed anomaly between time reversal and $\mathbb{Z}_2$. Notice that it implies that $\sf{T}$ exchanges the subspaces $\mathcal{H}_\pm$ with each other. In this situation, time reversal implies exact degeneracy between the two sectors, but doesn't place any restriction on either sector individually, so the ensemble consists of two degenerate GUE-like sectors.

\section{Summary and discussion}\label{sec:discussion}

In this paper, we studied two-dimensional gauge theory coupled to gravity as a dual description of certain matrix ensembles with global symmetry $G$.
First, we studied theories without anomalies.
The matrix ensemble splits into several smaller uncorrelated matrix ensembles labelled by the irreducible representations $r$ of $G$. The ground state energy in each sector is proportional to $c_2(r)$. The density of eigenvalues in each sector is proportional to $(\text{dim}\, r)^2$. One of these factors of $\text{dim}(r)$ arises from the exact degeneracy due to the $G$ symmetry. However, there is a leftover factor of $\dim(r)$, which implies that the number of copies of each irrep appearing in the spectrum is proportional to $\text{dim}(r)$. These copies are not exactly degenerate. One can speculate that they could be related to the existence of a second action of $G$ on the Hilbert space that commutes with the $G$-symmetry action but does not commute with the Hamiltonian. In other words, the second copy of $G$ acts in the space indexed by the letter $i$ in (\ref{iindex}), but does not commute with the Hamiltonian. This is reminiscent of the paper \cite{Lin:2019qwu} that constructed, in the setting of JT gravity coupled to matter, a gauge-invariant SL$_2$ algebra that does not commute with the Hamiltonian.\footnote{Juan Maldacena pointed out to us that the Hilbert space of an asymmetric top has this structure: the system has an $SO(3)$ symmetry, but there is a second copy of $SO(3)$ acting on the Hilbert space that does not commute with the Hamiltonian. See, for example, sections 45 and 46 of \cite{davydov}. One could also consider the bound state spectrum of the Hydrogen atom and consider perturbing away from a pure $1/r$ potential, which breaks the symmetry generated by the Runge-Lenz vector.} It may be important to understand this point better.

Next, we studied theories with anomalies.
The different sectors are now labelled by the projective representations of $G$.
We also studied $G$ symmetry in the presence of time reversal symmetry and concluded that self-conjugate representations correspond to GOE and GSE type blocks, whereas we get degenerate pairs of GUE blocks for representations with $r \neq \overline{r}$.
Finally, we studied a mixed anomaly between $\mathbb{Z}_2$ and time reversal, and concluded that that the Hilbert space has two sectors that are exchanged by the time reversal symmetry, and hence the matrix ensemble consists of two degenerate GUE-like sectors.

\subsection*{Acknowledgments}
We would like to thank Clay Cordova, Dan Freed, Luca Iliesiu, Alexei Kitaev, Juan Maldacena, Kantaro Ohmori, Shu-Heng Shao, Ryan Thorngren, Zhenbin Yang, Ying Zhao and especially Nathan Seiberg and Edward Witten  for helpful conversations.
DK gratefully acknowledges support from DOE grant DE-SC0009988 and the Chooljian family.
RM is supported by US Department of Energy grant No. DE-SC0016244.
The work of RM was performed in part at the Aspen Center for Physics, which is supported by National Science Foundation grant PHY-1607611.
RM also acknowledges fruitful conversations during the ``Developments in Quantum Field Theory and Condensed Matter" conference (SCGP, Stony Brook, Nov 5-7 2018) that was partially funded by the Simons Collaboration on the Non-perturbative Bootstrap.

\appendix

\section{Disk, trumpet, and double trumpet path integrals in the BF formulation}
\label{app:diskAndTrumpet}
In this appendix, we will compute some path integrals that are necessary for studying the BF theory on spaces with the ``non-topological'' type of boundary that we discussed in Section \ref{sec:BF} above. 
Our goal is to derive the formulas (\ref{disk}) and (\ref{trumpet}) for the disk and trumpet partition functions.
The action is given in (\ref{eq:fullaction}) supplemented with the boundary condition (\ref{eq:boundarycondition}).

\subsection{The disk}
\label{app:disk}
We start with the case of the disk. 
The path integral over $B$ in the interior of the disk imposes the constraint $F = 0$, so we have to integrate over flat connections. 
On the disk topology, a flat connection can be written as\footnote{We are suppressing a subtlety in the case of non-simply-connected groups that we will return to below.}
\begin{equation}
    A = g\mathrm{d} g^{-1}\label{ginvg},
\end{equation}
where the formal gauge transformation $g$ is a map from the disk to the group $G$. In the BF theory, we require true gauge transformations to vanish at the boundary, which means that the boundary values of $g$ cannot be changed by a gauge transformation. These boundary values are physical modes, and we need to path-integrate over them. On the boundary, (\ref{ginvg}) reduces to $ A_\tau = g\partial_\tau g^{-1} $, and the boundary action of the BF theory (\ref{eq:fullaction}) becomes (working with zero chemical potential at the moment)
\begin{equation}
    I = -\frac{1}{2}\int_0^\beta \mathrm{d}\tau \tr\left[\left(g\partial_\tau g^{-1}\right)^2\right].
\end{equation}
Here $g = g(\tau) \in G$ is a function of the time coordinate along the boundary. This action is the well-known 1d sigma model that describes a particle moving on the group manifold $G$.

There are subtly different path integrals that one can consider with this action. The most obvious option is to integrate $g(\tau)$ over the full space $\text{loop}(G)$ of maps from the circle to $G$, with the ultralocal measure that follows from the metric
\begin{equation}
    \|\delta g\|^2 = -\int_0^\beta \mathrm{d}\tau \tr\left[\left(g\delta g^{-1}\right)^2\right].\label{ultralocal}
\end{equation}
The path integral with the integration measure associated to this metric is most easily obtained using canonical quantization of the particle-on-a-group theory. 
The energy eigenfunctions of the corresponding quantum mechanical system are the matrix elements of $g$ in an irreducible representation $r$, namely $\psi_{r,ab}(g) = \langle r,a|g|r,b\rangle$, and the corresponding energies are $c_2(r)/2$. 
Since there are $\dim(r)^2$ such matrix elements, the degeneracy is $\text{dim}(r)^2$, and one finds the partition function \cite{marinov1979dynamics}
\begin{equation}
    Z_{\text{ultralocal}}(\beta) = \sum_{r}\text{dim}(r)^2 e^{-\beta c_2(r)/2}.\label{qmnorm}
\end{equation}

For the application to BF theory, this is not quite the right way to treat the path integral. One can see this by noting that under right-multiplication by a constant group element, $g(\tau)\rightarrow g(\tau)\cdot h$, the formula for $A$ in (\ref{ginvg}) is unchanged. This means that the space $\text{loop}(G)$ is a redundant description of the flat connections on the disk. Instead, we should integrate over the quotient of this space by right $G$ actions, denoted $ \text{loop}(G)/G$.

The appropriate measure for integrating over $\text{loop}(G)/G$ is the one that follows from the symplectic form of the BF theory:
\begin{equation}
    \omega(\delta_1A,\delta_2A) = 2\alpha\int_{\Sigma} \tr(\delta_1A\wedge \delta_2 A),\label{sympform}
\end{equation}
where $\delta_iA$ are small changes to the flat connection and $\Sigma$ is the spacetime manifold. 
We are interested in evaluating this symplectic form for configurations of the type (\ref{ginvg}), with variations parametrized by $\delta g$. It will be helpful to have a formula for the RHS of (\ref{sympform}) in the case where one of the variations is a gauge transformation, and the other variation $\eta$ preserves the flatness of $A$:
\begin{align}
    \int \tr\left[(\mathrm{d}\Theta + [A,\Theta])\wedge \eta\right]
    &= \int \tr\left[\mathrm{d}(\Theta \eta)-\Theta(\mathrm{d}\eta + \eta\wedge A +  A \wedge\eta)\right]\\
    &=\int_{\text{bdy}} \tr\left[\Theta\eta\right].\label{usedTo}
\end{align}
In the last step we used that $\mathrm{d}\eta + \eta \wedge A + A\wedge \eta = 0$ in order for the variation corresponding to $\eta$ to preserve the flatness of $A$. We can use (\ref{usedTo}) to compute $\omega(\delta_1 g,\delta_2g)$. 
We set $\Theta$ equal to the gauge transformation associated to $\delta_1 g$, which is $\Theta = g\delta_1 g^{-1}$, and $\eta$ equal to $\delta_2(g\mathrm{d}g^{-1})$. 
Simplifying slightly, one finds
\begin{equation}
    \omega(\delta_1 g,\delta_2 g) = 2\alpha\int_0^\beta \mathrm{d}\tau \tr\left[\left(\delta_1 g^{-1}\,g\right)\partial_\tau\left(\delta_2 g^{-1}\, g\right)\right] \, ,\label{sympformg}
\end{equation}
which can also be written as
\begin{equation}
    \omega = \alpha\int_0^\beta \mathrm{d}\tau\tr\left[\mathrm{d}g^{-1}\, g\wedge \partial_\tau(\mathrm{d}g^{-1}\, g)\right].\label{sympform3}
\end{equation}

It will be useful to compare the measure on $\text{loop}(G)/G$ induced by this symplectic form to the meaure on $\text{loop}(G)$ induced by the ultralocal measure (\ref{ultralocal}). 
As we will see, the ratio of these measures is a measure on the group $G$. 
First, we note that the symplectic form (\ref{sympform3}) is right-invariant under the action of $\text{loop}(G)$, and the ultralocal measure (\ref{ultralocal}) is invariant under both left and right actions, so it is sufficient to compare the measures at the identity point $g(\tau) = \text{id}$.
At the identity point, we can identify $\delta g^{-1}(\tau)$ with an element of the Lie algebra $t(\tau)\in \mathfrak{g}$, and we can decompose this function into Fourier modes
\begin{equation}
    t(\tau) = \sum_{n} e^{-2\pi \i n \frac{\tau}{\beta}}t_n^A T^A
    \label{eq:deftnA}
\end{equation}
where $T^A$ are anti-Hermitian and form an orthonormal basis for the Lie algebra of $G$, so that
\begin{equation}
    \tr T^AT^B = -\delta^{AB}.
\label{Tnorm}
\end{equation}
The set of Fourier modes parametrize the tangent space to $\text{loop}(G)$ at the identity, and the set of {\it nonzero} Fourier modes parametrize the tangent space to $ \text{loop}(G)/G$. So the ratio of the ultralocal and symplectic measures should be understood as a measure on the space of zero modes $t_0^A$.

To work out the precise measure, we write out the symplectic form explicitly in terms of the coordinates $t_n^A$. More precisely, we will write it in terms of the real and imaginary parts of $t_n^A$:
\begin{equation}
    \omega = -\alpha \sum_A\sum_{n\ge 1}(8\pi n)
    \, \mathrm{d}t_n^{A \,(R)}\wedge\mathrm{d}t_n^{A\,(I)}.
\end{equation}
The corresponding measure is
\begin{equation}
   \mu_{\text{symplectic}} =  \prod_A\prod_{n\ge 1}(8\pi \alpha n)\, \mathrm{d}t_n^{A\,(R)}\mathrm{d}t_n^{A\,(I)}.
\end{equation}
On the other hand, in the same coordinates the ultralocal measure is
\begin{equation}
  \mu_{\text{ultralocal}} = \prod_A \mathrm{d}t_0^{A} \prod_A\prod_{n\ge 1}2\mathrm{d}t_n^{A\,(R)}\mathrm{d}t_n^{A\,(I)}.\label{nbm}
\end{equation}
Note that we have kept careful track of the overall multiple $\alpha$ of the symplectic form, but we have not kept track of a similar constant multiple of the metric (\ref{ultralocal}). 
In fact, in writing (\ref{nbm}), we implicitly adjusted the overall normalization of (\ref{ultralocal}) in order to cancel some factors.
The reason is that the measure associated to (\ref{ultralocal}) is actually independent of rescalings of the metric (\ref{ultralocal}). 
Rescaling the metric would multiply the measure by factors of the form $\prod_{n} (\text{const.})$, which is equal to one after renormalization. On the other hand, multiplying $\omega$ by a constant would rescale the symplectic measure by a factors of the form $\prod_{n\neq 0}(\text{const.})$, which is not equal to one.

With this understanding, we can now compute the ratio
\begin{equation}
    \frac{\mu_{\text{ultralocal}}}{\mu_{\text{symplectic}}} = \left(\prod_{n\ge 1}\frac{1}{4\pi \alpha n}\right)^{\text{dim}(G)}\prod_A \mathrm{d}t_0^{A}= (2\alpha)^{\frac{1}{2}\text{dim}(G)}\prod_A \mathrm{d}t_0^A,
    \label{eq:muovermu}
\end{equation}
where in the final equality we took the renormalized product (e.g.~using zeta function regularization). 
The final expression in (\ref{eq:muovermu}) is a measure on the Lie algebra of the group, normalized according to the metric
\begin{equation}
    \langle t,s\rangle = - 2\alpha\tr(ts)\, ,
    \label{metLie}
\end{equation}
where $t,s \in \mathfrak{g}$. This metric on the Lie algebra induces a bi-invariant metric on the group, 
\begin{equation}
    \mathrm{d}s^2= - 2\alpha \tr(g\mathrm{d}g^{-1})^2
    \label{eq:metLie2}
\end{equation}
and we denote the corresponding volume as $\vol(G)$. Note that this volume depends on the normalization coefficient $\alpha$. For example, $\vol(SU(2))= (4\alpha)^\frac{3}{2} \, 2\pi^2$.

 Our conclusion is that the path integral over $\text{loop}(G)/G$ with the symplectic measure will be related to the path integral over $\text{loop}(G)$ with the ultralocal measure according to 
\begin{equation}
   Z_{\text{loop}(G)/G}(\beta)= \frac{1}{\vol(G)}\, Z_{\text{ultralocal}}(\beta).
   \label{eq:z-loopg-mod-g}
\end{equation}
The expression on the LHS is the sum over all representations of the function $Z_D^{\text{gauge}}(\beta,r)$ in (\ref{disk}). In section \ref{sec:chemicalApp} we will show how to get the formula for fixed representation.

We will now address a subtlety that arises in the case of non-simply connected groups. Suppose that we have a gauge field of the form (\ref{ginvg}). In order for this to be nonsingular everywhere in the interior of the disk, the image of the boundary of the disk under $g$ must be a contractible path in the group $G$. However, for non-simply-connected groups, the integral over $\text{loop}(G)$ or $\text{loop}(G)/G$ includes configurations $g(\tau)$ that do not have this property.

Such configurations can be incorporated in the BF theory by including nontrivial bundles for the bulk gauge field $A$.\footnote{Normally, all bundles over the disk are trivial. But with fixed gauge at the boundary, there are nontrivial bundles.} On the disk topology, these bundles can be described using two patches. For example, if we use a radial coordinate $r$ that runs between zero and one, then we can use a patch near the center $P_1 = \{r<2/3\}$ and an annular region that includes the boundary, $P_2 = \{1/3 < r \le 1\}$. The intersection is an annular region $\{1/3 < r < 2/3\}$, and the bundle is described by a transition function that maps this region into $G$. By using a transition function that winds around the desired noncontractible cycle of $G$ as the angular coordinate winds around the annulus, we can find a nonsingular flat connection that has the form (\ref{ginvg}) in the outer patch, with noncontractible $g(\tau)$.

For example, in the case of $G = U(1)$, we can take
\begin{equation}
    A\big|_{P_1} = 0, \hspace{20pt} A\big|_{P_2} = -\mathrm{i} n\,\mathrm{d}\theta.
\end{equation}
This provides a flat connection on the bundle defined by the transition function 
\begin{equation}
    g_{12}:P_1\cap P_2 \rightarrow U(1), \hspace{20pt} g_{12}(r,\theta) = e^{\mathrm{i}n\theta}.
\end{equation}
In the outer region $P_2$, the gauge field is equal to (\ref{ginvg}) with a $g$ that winds $n$ times around $U(1)$.

\subsection{The trumpet}
Another useful path integral is the ``trumpet'' path integral. This space is topologically a cylinder, but we will use the term trumpet to emphasize that the two boundaries are treated differently. In particular, at the ``big end'' of the trumpet, we have a non-topological boundary condition that supports the particle-on-a-group theory. The other ``topological" end will always be glued onto another topological boundary, and we will work out how this gluing works later.

For the moment, we are interested in the path integral of the particle-on-a-group theory on the non-topological boundary. The analysis of this system is similar to that of the disk path integral considered above, with one important distinction: because the cylinder contains a non-contractible cycle, there exist additional non-trivial flat connections with holonomy $U$.

For example, if we take the periodic direction of the cylinder to be parametrized by $\tau$, then a gauge field 
\begin{equation}
    A = a\mathrm{d}\tau
\end{equation}
would describe a situation with holonomy $U = e^{-\beta a}$. A more general flat connection would be a formal gauge transformation of this one,
\begin{align}
    A = g (a \d \tau) g^{-1} + g\mathrm{d}g^{-1}.\label{param2}
\end{align}
The particle-on-a-group theory for the trumpet arises from the path integral over the boundary values of $g(\tau)$ on the boundary at ``big end'' of the trumpet.

To analyze this theory, it is convenient to change variables to $\widetilde{g}(\tau) = g(\tau) e^{-a\tau}$. Then we have simply
\begin{equation}
    A =  \widetilde{g}\mathrm{d}\widetilde{g}^{-1}.
\end{equation}
However, note that $\widetilde{g}$ is no longer periodic, but instead satisfies $\widetilde{g}(\beta) =  \widetilde{g}(0)U$, where $U = e^{-\beta a}$. It makes sense to consider the particle-on-a-group theory with this non-periodic boundary condition and with the ultralocal measure associated to (\ref{ultralocal}). Using the Hilbert space for the particle on the group manifold, one finds (see section \ref{sec:chemicalApp} below for some details on a slightly more general computation) 
\begin{equation}
    Z_{\text{ultralocal}}(\beta;U) = \sum_r \text{dim}(r)\chi_r(U)e^{-\beta c_2(r)/2},\label{ultralocalTrumpet}
\end{equation}
which reduces to the disk computation (\ref{qmnorm}) for $U = 1$. 

The path integral we actually want is the one with the symplectic measure (\ref{sympform}). To work this out explicitly, we start with the expression for $A$ in (\ref{param2}). The value of $g$ in the bulk is not meaningful since it can be changed by a true gauge transformation that vanishes on the boundary, but the values of $g$ on the big end of the trumpet are physical. Even so, the parametrization (\ref{param2}) is slightly redundant, since multiplication $g(\tau)\rightarrow g(\tau)\cdot h$ by a group element $h$ that commutes with $U$ preserves the form of $A$. 
For a generic $U$, the subgroup of $G$ that commutes with $U$ will be a maximal torus of the Lie group, which we refer to as $T$. 
This redundancy is reflected in the fact that, for a fixed $a$, the symplectic form \eqref{sympform} is degenerate and its zero modes span the maximal torus. 
To obtain a non-zero answer using the symplectic measure, we must therefore quotient the domain of integration in the path integral by the right-action of elements of $T$. 
In other words, we should integrate the boundary values of $g$ over the space $\text{loop}(G)/T$.

We are interested in the path integral over $\text{loop}(G)/T$, with the measure that follows from the symplectic form \eqref{sympform}.
The induced measure on the space of $g$ fluctuations, generalizing \eqref{sympform3}, is
\begin{equation}
    \omega = \alpha \int_0^\beta \mathrm{d}\tau \tr\left[(\mathrm{d} g^{-1}\,g)\wedge D_\tau(\mathrm{d} g^{-1}\,g)\right], \hspace{20pt} D_\tau = \partial_\tau + [a,\cdot].\label{sympform4}
\end{equation}
We would like to compare the measure associated to this symplectic form with the ultralocal measure. 
Because of right-invariance, it is enough to work out the ratio of these measures close to the identity, where we can view $\mathrm{d}g$ as a Lie algebra element. 
We decompose the Lie algebra $\mathfrak{g} = \mathfrak{g}_0\oplus \mathfrak{g}^\perp$, where $\mathfrak{g}_0$ is the Lie algebra of the maximal torus $T\subset G$ that contains $U$, and $\mathfrak{g}^\perp$ is the orthocomplement of $T$ in $G$. 
Then, for the fluctuation components in $\mathfrak{g}_0$, the calculation is the same as for the disk computation above. 
In particular, the ratio of the ultralocal and symplectic measures in the $\mathfrak{g}_0$ directions is a measure on the space of zero modes in $\mathfrak{g}_0$ itself.

For the fluctuations in the $\mathfrak{g}^\perp$ directions, we have to do a new calculation. In general, the measure associated to a symplectic form $\omega = \frac{1}{2}\omega_{ij}\mathrm{d}x^i \mathrm{d}x^j$ is given by 
\begin{align}
  \mu &= \frac{\omega^n}{n!} = \mathrm{d}x^1\dots \mathrm{d}x^{2n} \,\text{Pf}(\omega_{ij}) \\
  &=\mathrm{d}x^1\dots \mathrm{d}x^{2n}\int \mathrm{d}\psi^1\dots \mathrm{d}\psi^{2n} \exp\left(\frac{1}{2}\omega_{ij}\psi^i\psi^j\right).
\end{align}
In the final expression, the $\psi^i$ are Grassman variables. 
To compute the Pfaffian of the symplectic form $\omega$, we replace $\mathrm{d}x^i\rightarrow \psi^i$ in the symplectic form and integrate $\exp \omega$ over the variables $\psi^i$.
For the infinite dimensional case we are interested in, this becomes a Grassman path integral.

It will be enough to consider the case of $\mathfrak{g} = \text{su}(2)$. We will work with the basis for the Lie algebra
\begin{equation}
    X = \left(\begin{array}{cc} 0 & 1 \\ 1 & 0 \end{array}\right),\hspace{20pt}Y= \left(\begin{array}{cc} 0 & -\mathrm{i} \\ \mathrm{i} & 0 \end{array}\right),\hspace{20pt}Z = \left(\begin{array}{cc} 1 & 0 \\ 0 & -1 \end{array}\right).
\end{equation}
This is a different normalization than (\ref{Tnorm}), but as we will see below, this can be absorbed into the constant $\alpha$ below.
We take $a = \mathrm{i}a_Z Z$ and we want to compute the Pfaffian of the symplectic form for the components corresponding to fluctuations in the $X$ and $Y$ Lie algebra directions, working close to the identity. To do so, we make the replacement in the symplectic form (\ref{sympform4})
\begin{equation}
  \mathrm{d}g^{-1}(\tau)\, g(\tau)\rightarrow \mathrm{i}\psi_X(\tau) X + \mathrm{i}\psi_Y(\tau) Y
\end{equation}
and then evaluate the path integral over $\psi_X(\tau)$ and $\psi_Y(\tau)$. This is
\begin{align}
    \text{Pf}_\perp(\omega)&=\int \mathcal{D}\psi_X\mathcal{D}\psi_Y \exp\left(-\alpha\int_0^\beta \mathrm{d}\tau\tr\left[(\psi_XX  + \psi_YY)(\psi'_XX + \psi'_Y Y - 2a_Z (\psi_X Y - \psi_YX)\right]\right)\notag\\
    &=  \int \mathcal{D}\psi_X\mathcal{D}\psi_Y \exp\left(-2\alpha\int_0^\beta \mathrm{d}\tau\left[\psi_X\psi'_X + \psi_Y \psi'_Y +  4a_Z\psi_X\psi_Y\right]\right).\label{pathIntIs}
\end{align}
Here  $\psi'$ denotes a derivative of $\psi$ with respect to $\tau$. Because we are integrating over all modes, the answer is independent of $\alpha$ and choosing $\alpha = \frac{1}{4}$, this becomes a more-or-less conventional path integral, with two Majorana fermions satisfying periodic boundary conditions in Euclidean time. 
The path integral can be done by performing an explicit mode decomposition, but it is evaluated most easily by quantizing the theory in a two-dimensional Hilbert space, e.g.
\begin{equation}
    \psi_X = \frac{1}{\sqrt{2}}X, \hspace{20pt} \psi_Y = \frac{1}{\sqrt{2}}Y.
\end{equation}
The Hamiltonian is $H = 2a_Z\psi_X\psi_Y = \mathrm{i}a_Z Z$. The path integral (\ref{pathIntIs}) is
\begin{equation}
    \text{Pf}_\perp(\omega) = \tr \i (-1)^F e^{-\beta H} = 2\sin(\beta a_Z)
\end{equation}
where we inserted $\mathrm{i}(-1)^F = \mathrm{i}Z$ to impose periodic boundary conditions for $\psi_X$ and $\psi_Y$.

In order to generalize this computation beyond $\text{su}(2)$, it is helpful to rewrite the answer in terms of $\det_{\mathfrak{g}^\perp}(1-U)$ where $U$ acts on $\mathfrak{g}^\perp$ via the adjoint action. Continuing with our $SU(2)$ example, we have $U = e^{-\mathrm{i}\beta a_Z Z}$. The eigenvalues of $Z$, acting in the adjoint representation, are $\pm 2$. So 
\begin{equation}
    \text{det}_{\mathfrak{g}^\perp}(1-U) = (1-e^{2\mathrm{i}\beta a_Z})(1-e^{-2\mathrm{i}\beta a_Z}) =  \left[2\sin(\beta a_Z)\right]^2\, .
\end{equation}
This implies that for the case of $\text{su}(2)$, we have
\begin{equation}
    \text{Pf}_\perp(\omega) = \sqrt{\text{det}_{\mathfrak{g}^\perp}(1-U)}.\label{detAns}
\end{equation}
For a general Lie algebra, we will have a copy of the $\text{su}(2)$ calculation for each pair of Lie algebra generators that correspond to a given root. Taking the product over all such factors, one finds that (\ref{detAns}) is actually the general formula.

So the path integral on $ \text{loop}(G)/T$ is
\begin{align}
    Z_{ \text{loop}(G)/T}(\beta;U) &= \frac{\sqrt{\det_{\mathfrak{g}^\perp}(1-U)}}{\vol(T)}
    Z_{\text{ultralocal}}(\beta;U)\, ,
    \label{eq:nonabeliantrumpet}
\end{align}
where $Z_{\text{ultralocal}}(\beta; U)$ was given in (\ref{ultralocalTrumpet}).
The factor in the numerator is (\ref{detAns}), and the factor in the denominator arises from the integral over fluctuations in the $\mathfrak{g}_0$ directions, as explained before.

\subsection{The double trumpet}
\label{app:doubletrumpet}
The double trumpet is a space with the topology of the cylinder, with the non-topological boundary condition (\ref{eq:boundarycondition}) on both ends, each supporting a particle-on-a-group mode.
The path integral on this space is important, for its own sake. 
However, it is also useful because the double trumpet can be constructed by gluing two ordinary trumpets together along their ``small ends.'' 
So comparison of the trumpet path integral and the double trumpet path integral will teach us the gluing rule for the small end of a trumpet.

On the cylinder topology, a flat gauge field can be parametrized as\footnote{For non-simply-connected $G$, this is too restrictive, since it implies that the $g(\tau)$ functions on the left and right boundaries are homotopic. This restriction can be lifted by using nontrivial bundles as described for the disk computation above.}
\begin{equation}
    A = g (a \mathrm{d}\tau) g^{-1} + g\mathrm{d}g^{-1} \, ,
\end{equation}
just as in (\ref{param2}).
The path integral for the double trumpet involves an integral over the particle-on-a-group modes associated to both boundaries, together with an integral over the holonomy parametrized by $a$. The measure for all of these pieces is induced by the symplectic form (\ref{sympform}).

The new component of the symplectic form that we have to evaluate is $\omega(\delta g,\delta a)$. 
To evaluate this, it is convenient to use (\ref{usedTo}) with $\Theta$ a gauge transformation corresponding to $\delta g$, namely $\Theta = g\delta g^{-1}$. 
The second parameter $\eta$ corresponds to the variation of $a$, so $\eta = g (\delta a\, \mathrm{d}\tau) g^{-1}$.
We get
\begin{equation}
    \omega(\delta g,\delta a) = 2\alpha \int_0^\beta \mathrm{d}\tau \Tr\left[(\delta g_R^{-1} g_R - \delta g_L^{-1} g_L )\delta a\right].
\end{equation}
Here $g_L$ and $g_R$ are the values of $g$ on the two boundaries of the double trumpet. We can therefore write the full symplectic form as
\begin{equation}
    \omega = \alpha \int_0^\beta \mathrm{d}\tau\tr\left[(\mathrm{d} g_R^{-1}\,g_R)\wedge D_\tau(\mathrm{d} g_R^{-1}\,g_R) + 2\mathrm{d}g_R^{-1}\,g_R\wedge \mathrm{d}a\right] - (R\rightarrow L).
\end{equation}
In this expression, $g_L$ and $g_R$ are functions of $\tau$, but $a$ is constant.

As before, we would like to work out the measure induced by this symplectic form. 
By right-invariance, it is sufficient to work near the identity for $g_L$ and $g_R$, so that $\mathrm{d}g_L^{-1} = \mathrm{d}t_L$ and $\d g_R^{-1} = \d t_R$ are Lie algebra elements.  
For the nonzero Fourier modes of $\mathrm{d}g^{-1}$, the last term involving $a$ drops out, and we find the same symplectic form as for the trumpet calculation. For the Fourier zero modes, however, the situation is different. Writing the Fourier zero modes as $t_{0\, R}$ and $t_{0\, L}$, it is convenient to define
\begin{equation}
    t_\pm = t_{0\, R} \pm t_{0\, L}.
\end{equation}
Then the symplectic form can be written as
\begin{equation}
    \omega\big|_{\text{zero modes}} = \alpha\beta \tr\Big[\mathrm{d}t_-\wedge\big(2\mathrm{d}a + [a,\mathrm{d}t_+]\big)\Big].\label{sympform5}
\end{equation}
This form appears to be degenerate, because there are linear combinations of $\mathrm{d}t_+$ and $\mathrm{d}a$ that are zero modes of $\omega$. 
However, all this means is that our coordinates on the space of gauge fields are overcomplete. 
A quick way to identify a minimal set is to restrict to the set of nonzero modes of the symplectic form. 
A convenient restriction is to 
\begin{equation}
    t_- \in \mathfrak{g}, \hspace{20pt} t_+ \in \mathfrak{g}^\perp, \hspace{20pt} a \in \mathfrak{g}_0.
\end{equation}
When $t_-\in \mathfrak{g}^\perp$, the term proportional to $2\mathrm{d}a$ in \eqref{sympform5} vanishes. The part of the integral with $t_-$ in $\mathfrak{g}^\perp$ therefore combines with the integral with $t_+$ in $\mathfrak{g}^\perp$ to give the same zero mode integral that is found for the separate computation of two trumpets, as discussed in the last subsection. 

It remains to understand the measure for the integral over $t_-$ and $a$ in $\mathfrak{g}_0$. For these modes, the term proportional to the commutator in \eqref{sympform5} vanishes. Since $\beta \mathrm{d}a = \mathrm{d}U\,U^{-1}$, the Pfaffian of the restriction of (\ref{sympform5}) to these modes is just the measure associated to the Lie algebra metric (\ref{metLie}), restricted to $\mathfrak{g}_0$. Thus the double trumpet with no restriction on the representations on the two ends is given by
\begin{align}
    Z_{0}^{\text{gauge}}(\beta_1,\beta_2) &= \vol(T)\int_{T} \mathrm{d}U Z_{\text{loop}(G)/T}(\beta_1,U)Z_{\text{loop}(G)/T}(\beta_2,U^{-1})\, .\label{secondfactor}
    \end{align}
Here, the factor of $\vol(T)$ came from the integral over the component of $t_-$ in $\mathfrak{g}_0$, extended appropriately away from the identity. The integral $\mathrm{d}U$ is over the torus, with the measure associated to the Lie algebra metric (\ref{metLie}).\footnote{In the second factor of (\ref{secondfactor}), we have replaced $U$ by $U^{-1}$ because we are measuring the holonomy relative to an orientation of the boundary inherited from the bulk, and this orientation corresponds to opposite direction around the cylinder at the two ends.} Plugging in (\ref{eq:nonabeliantrumpet}) and (\ref{ultralocalTrumpet}), the RHS is explicitly
\begin{align}
  \int_{T}\frac{\mathrm{d}U}{\vol(T)}\text{det}_{\mathfrak{g}^\perp}(1-U) \sum_{r,r'}\dim(r)\dim(r')\chi_r(U)\chi_{r'}(U^{-1})e^{-\beta_1c_2(r)/2-\beta_2 c_2(r')/2}.\label{togen}
  \end{align}
  Using the Weyl integral formula, the integral over the torus $T$ can be replaced by an integral over the full group $G$:
\begin{align}
    \int_{G}\frac{\mathrm{d}U}{\vol(G)}\sum_{r,r'}\dim(r)\dim(r')\chi_r(U)\chi_{r'}(U^{-1})e^{-\beta_1c_2(r)/2-\beta_2 c_2(r')/2}.
    \end{align}
    Finally, using the orthonormality of the characters, we find the simple expression 
    \begin{equation}
      Z_{0}^{\text{gauge}}(\beta_1,\beta_2) =  \sum_{r} \dim(r)^2 e^{-(\beta_1 + \beta_2)c_2(r)/2}.\label{doubletrumpetans}
    \end{equation}
Our route to this formula was somewhat involved. It seems likely that there is a quicker method, using quantization of the BF theory on the interval that stretches between the two ends of the trumpets. Indeed, the fact that the coefficients that appear in (\ref{doubletrumpetans}) must be integers follows immediately from that perspective.

\subsection{Including a chemical potential}
\label{sec:chemicalApp}
We would like to work out the formula for the trumpet or disk partition function when we include a chemical potential $\mu$. 
The action for the BF theory including $\mu$ was given in (\ref{eq:fullaction}), with the boundary condition in (\ref{eq:boundarycondition}). 
The corresponding action for the particle-on-a-group boundary mode is obtained by substituting (\ref{param2}) into the boundary action
:
\begin{equation}
    I = -\frac{1}{2}\int_0^\beta \mathrm{d}\tau \tr\left[(g\partial_\tau g^{-1} + ga g^{-1} + \mu)^2\right].\label{generator}
\end{equation}
This formula is appropriate for the disk when $a = 0$ and appropriate for the trumpet when $a\neq 0$. The global $G$ symmetry for which $\mu$ is a chemical potential corresponds to left-multiplication
\begin{equation}
    g(\tau)\rightarrow h\cdot g(\tau).
\end{equation}
With a generic chemical potential included, the symmetry is reduced from $G$ to a maximal torus $T$.

We would like to understand how the above computations of the disk and trumpet path integrals are modified by the $\mu$ term. It is convenient to rewrite the action (\ref{generator}) as 
\begin{equation}
    I = -\frac{1}{2}\int_0^\beta \mathrm{d}\tau \tr\left[(\widetilde{g}\partial_\tau\widetilde{g}^{-1})^2\right], \hspace{20pt} \widetilde{g}(\tau) = e^{-\mu\tau}g(\tau)e^{-a\tau}.
\end{equation}
So we have the ordinary particle-on-a-group action, but with a variable $\widetilde{g}$ that is not periodic in time, and instead satisfies  $\widetilde{g}(\beta) = V^{-1} \widetilde{g}(0)U$. Here $U = e^{-\beta a}$ and $V = e^{\beta \mu}$.

With respect to the ultralocal measure, this path integral can be done by quantizing the particle-on-a-group theory. The eigenfunctions of the Hamiltonian are the matrix elements of $g$ in a given representation $\psi_{r,ab}(g) = \langle r,a|g|r,b\rangle$, and the corresponding energy eigenvalue is $c_2(r)/2$. 
This gives
\begin{align}
Z_{\text{ultralocal}}(\beta,V;U) &= 
\sum_{rab}\int \mathrm{d}g\;\psi_{r,ab}(V^{-1}gU)\overline{\psi_{r,ab}(g)}e^{-\beta c_2(r)/2} \\
&=\sum_{r}\chi_r(V^{-1})\chi_r(U)e^{-\beta c_2(r)/2}\, ,
\label{eq:zultrauv}
\end{align}
which generalizes the earlier formulas (\ref{ultralocalTrumpet}) and (\ref{qmnorm}).

We can use (\ref{projectOnto}) to find a corresponding formula with fixed representation instead of fixed chemical potential:
\begin{equation}
    Z_{\text{ultralocal}}(\beta,r;U)= \dim(r)\chi_r(U) e^{-\beta c_2(r)/2}.
\end{equation}
Since we didn't change the measure when we introduced the chemical potential, the relationship between the ultralocal and symplectic path integrals remains the same, and in particular from (\ref{eq:z-loopg-mod-g}) and (\ref{eq:nonabeliantrumpet}), we have
\begin{align}
    Z_{\text{loop}(G)/G}(\beta,r) &= \frac{\dim(r)^2}{\vol(G)}e^{-\beta c_2(r)/2},\label{diskdisk}\\
     Z_{\text{loop}(G)/T}(\beta,r;U) &=\dim(r) \frac{\sqrt{\text{det}_{\mathfrak{g}^\perp}(1-U)}}{\vol(T)}\chi_r(U)e^{-\beta c_2(r)/2}.\label{trumpettrumpet}
\end{align}
Eq.~(\ref{diskdisk}) is the formula reported in (\ref{disk}).

To derive the formula for the trumpet (\ref{trumpet}) from (\ref{trumpettrumpet}), we need to transform the boundary condition at the ``small end'' to the boundary condition that can be glued using the gluing rule (\ref{gluingrule}). 
We will do this using an indirect argument. 
First, the corresponding formula for the double trumpet with fixed representation boundary conditions is
\begin{align}
    Z_0^{\text{gauge}}(\beta_1,r_1;\beta_2,r_2) &=
  \int_{T}\frac{\mathrm{d}U}{\vol(T)}\text{det}_{\mathfrak{g}^\perp}(1-U) \dim(r_1)\dim(r_2)\chi_{r_1}(U)\chi_{r_2}(U^{-1})e^{-\beta_1c_2(r_1)/2-\beta_2 c_2(r_2)/2}\notag\\
  &= \dim(r_1)^2\delta_{r_1,r_2}e^{-(\beta_1+\beta_2)c_2(r_1)/2}.\label{doubletrumpetrep}
\end{align}
Now, if $Z_T^{\text{gauge}}(\beta,r;r')$ is the trumpet with ``gluable'' boundary conditions for $r'$, then we should have 
\begin{equation}
    Z_{0}^{\text{gauge}}(\beta_1,r_1;\beta_2,r_2) =  \sum_{r'}Z^{\text{gauge}}_{T}(\beta_1,r_1;r')Z^{\text{gauge}}_{T}(\beta_2,r_2;r').
\end{equation}
Comparing to (\ref{doubletrumpetrep}), we find 
\begin{equation}
    Z_T^{\text{gauge}}(\beta,r;r') = \delta_{r,r'}\dim(r) e^{-\beta c_2(r)/2}.
\end{equation}
This is the result that was reported in (\ref{trumpet}).

\section{The case of U(1) symmetry}\label{sec:Abelian}

In this appendix, we write down the partition functions on a general Riemann surface for the special case of $U(1)$ symmetry.
The motivation for this case is the complex SYK model, see \cite{Gu:2019jub} for recent work. 
Other references include \cite{Fu:2016vas, Sachdev:2015efa}.
We can take all the results directly from Appendix \ref{app:diskAndTrumpet}, but since the $U(1)$ case is considerably simpler, and of some interest in its own right, we present it here separately, in brief.
The action is given in (\ref{eq:fullaction}), with boundary condition (\ref{eq:boundarycondition}).

The path integral on the disk can be reduced to a path integral over the space $\text{loop}(G)/G$.
The space of flat connections can be parametrized as $A = \mathrm{i}\,\d \Theta$, for which the symplectic form (\ref{sympform}) evaluates to
\begin{align}
    \omega(\delta_1 A, \delta_2 A) &= 
    -2\alpha \int_\Sigma \d\delta_1\Theta \wedge \d \delta_2\Theta
    = -2\alpha \int_{\partial\Sigma} \delta_1\Theta\, 
    \d \delta_2\Theta\, .
\end{align}
Now we parametrize 
$\Theta_{\text{bdy}}(\tau) = \sum_{m \neq 0} \, (t_m^{(R)} + \i t_m^{(I)})\, e^{- \i 2\pi m \frac{\tau}{\beta}} $ as in (\ref{eq:deftnA}), use the symplectic form to find the path integral measure, and do the Gaussian integrals.
The one-loop determinant equals
\begin{align}
    \prod_{m\geq 1} (2\alpha) (2\pi m) (2) \, \frac{\pi}{4\pi^2 m^2/\beta}
    &= \prod_{m \geq 1} \frac{2\alpha\beta}{m} =\left(\frac{1}{4\pi\alpha\beta} \right)^\frac{1}{2}\, .
\end{align}
Since $U(1)$ is not simply connected, we have to include the winding configurations discussed at the end of Section \ref{app:disk}.
The essential point is that a gauge field configuration can have unit holonomy with multiple possible configurations of $\Theta = 2\pi n\frac{\tau}{\beta}$.
Summing over different saddles, we get the final answer:
\begin{align}
    Z_D^{\text{gauge}}(\beta, \mu) =
    \left(\frac{1}{4\pi \alpha\beta} \right)^\frac{1}{2} 
    \sum_{n \in \mathbb{Z}} \exp \left[ 
    -\frac{1}{2\beta} \left( 2\pi n - \i \mu \beta \right)^2
    \right]\, .
    \label{eq:zdisk-u-one}
\end{align}
We can do a Poisson resummation and rewrite the result (\ref{eq:zdisk-u-one}) as
\begin{align}
    Z_D^{\text{gauge}}(\beta, \mu) = \frac{1}{(8\pi^2 \alpha)^\frac{1}{2}} \sum_{q \in \mathbb{Z}}
    \exp \left[q \, \beta\, \mu \right]\,
    \exp \left[ - \beta \, \frac{q^2}{2} 
    \right]
    \label{eq:zdisk-u-one-m}\, ,
\end{align}
which is a special case of (\ref{eq:z-loopg-mod-g}) and (\ref{eq:zultrauv}) with $U=1$ and $V= e^{\beta \mu}$.
(The volume of the group $U(1)$ evaluated with the metric (\ref{eq:metLie2}) on the group is $(2\alpha)^\frac{1}{2}2\pi$.)
We interpret the right hand side of (\ref{eq:zdisk-u-one-m}) as a sum over the $U(1)$ charges.
The overall constant means that we renormalize $e^{S_0}$ as
$e^{S_0} \to e^{S_0}(8\pi^2 \alpha)^{-\frac{1}{2}}$.
We can Laplace transform this expression to obtain a density of states, which will take the general form (\ref{eq:rho0rx}).
We emphasize that the path integral (\ref{eq:zdisk-u-one-m}) does not have a Hilbert space interpretation, and is subtly different from the path integral of a particle-on-a-group, because the zero mode is not being integrated over.

The partition function on the trumpet geometry reduces to a path integral on $\text{loop}(G)/T$, but since $T=G$ for $U(1)$, we will get the same expression as on the right hand sides of (\ref{eq:zdisk-u-one}) and (\ref{eq:zdisk-u-one-m}).
The difference is that if there is a holonomy $U = e^{\mathrm{i}\varphi}$ on the inner geodesic boundary of the trumpet, then $\beta\mu$ will be shifted to $\beta\mu - \mathrm{i}\varphi$.

As an example of a gluing calculation, we discuss the partition function for a double trumpet, which was presented for a general group in Section \ref{app:doubletrumpet}.
There is considerable simplification since the ``maximal torus" is the entire group as $g^\perp$ is empty.
The factor of $\vol(T)$ and the integral over $U$ in (\ref{secondfactor}) are simply interpreted as integrals over the holonomy from the one end of the double trumpet to the other, and the holonomy along the waist, respectively.
The result is (\ref{doubletrumpetans}).

For a general orientable Riemann surface, we find
\begin{align}
    Z_{g}^{\text{gauge}}(\beta_1, \mu_1; \ldots; \beta_n, \mu_n) = 
    (8\pi^2 \alpha)^\frac{2g+n-2}{2}
    \sum_{q \in \mathbb{Z}} 
    e^{q \beta (\mu_1 + \ldots + \mu_n)}\,
    e^ {- (\beta_1 + \ldots + \beta_n) \, \frac{q^2}{2}} \, .
\end{align}

\section{AdS-JT gravity plus Yang-Mills theory in the bulk}
\label{sec:yangmills}
Throughout this paper we have considered either the Yang-Mills theory with a coupling constant that is delta-function peaked on the boundary, or the BF theory with a special non-topological boundary condition. Either way, the theory is topological in the bulk, and this allowed us to be agnostic as to what the gravity theory is, since the full partition function factorized between the gravity sector and the gauge sector (on a given topology).

However, one could also contemplate non-topological gauge theories. A mild version of this is the standard Yang-Mills theory in the bulk, with a uniform coupling constant $e^2$. In this case, in a given representation sector, the gauge theory path integral depends on the metric, but only in a very simple way \cite{Witten:1991we}
\begin{align}
    Z_{g}^{YM}(U_1; \ldots; U_n) &\propto \sum_r 
    (\dim r)^\chi \chi_{r}(U_1) \ldots \chi_r(U_n) \exp 
    \left[-e^2 A c_2(r)/2\right]\, ,\label{projecting}
\end{align}
where $A$ is the area of the surface. We would like to make two comments about this.

First, the action of JT gravity can be written as a linear combination of the Euler characteristic, of $A$ and of $L$ where $L$ is the boundary length \cite{Kitaev:2018wpr,Yang:2018gdb}. So in a given representation sector, the effect of the Yang-Mills theory is just to change the coefficient of the $A$ term. This means that JT gravity coupled to Yang-Mills (even with finite coupling) is as solvable as ordinary JT gravity.

Second, in the case where we define JT gravity with asymptotic boundaries rather than boundaries of finite length, the area $A$ will go to infinity. In order to avoid projecting onto a single representation in (\ref{projecting}), we will need to take $e^2\rightarrow 0$ in this limit, in such a way that $e^2 A$ is finite. Because we are working in hyperbolic space, the area $A$ will be proportional to the boundary length, and in fact we will find
\begin{equation}
    e^2 A \rightarrow \# (\beta_1+\dots+\beta_n)
\end{equation}
where $\beta_j$ are the renormalized lengths of the boundary components, and $\#$ is a constant that can be absorbed into a redefinition of $\beta$. In the limit of asymptotic boundaries, then, JT gravity coupled to Yang-Mills theory reduces to the rather odd theory we studied in the main text, with $e^2$ concentrated in a delta function at the boundaries.

\bibliographystyle{apsrev4-1long}
\bibliography{JTYM2}
\end{document}